\documentclass{pasj01}
\draft 
\Received{$\langle$28-Feb-2019$\rangle$}
\Accepted{$\langle$25-Sep-2019$\rangle$}
\Published{$\langle$publication date$\rangle$}
\begin{document}

\title{G-0.02-0.07, the Compact H$_{\mathrm{II}}$  Region Complex nearest to the Galactic Center with ALMA}
\author{Masato Tsuboi$^{1, 2}$, Yoshimi Kitamura$^1$,  Kenta Uehara$^2$, Atsushi Miyazaki$^3$, Ryosuke Miyawaki$^4$, Takahiro Tsutsumi$^5$,  and Makoto Miyoshi$^6$}%
\altaffiltext{1}{Institute of Space and Astronautical Science, Japan Aerospace Exploration Agency,\\
3-1-1 Yoshinodai, Chuo-ku, Sagamihara, Kanagawa 252-5210, Japan }
\email{tsuboi@vsop.isas.jaxa.jp}
\altaffiltext{2}{Department of Astronomy, The University of Tokyo, Bunkyo, Tokyo 113-0033, Japan}
\altaffiltext{3}{Japan Space Forum, Kanda-surugadai, Chiyoda-ku,Tokyo,101-0062, Japan}
\altaffiltext{4}{College of Arts and Sciences, J.F. Oberlin University, Machida, Tokyo 194-0294, Japan}
\altaffiltext{5}{National Radio Astronomy Observatory,  Socorro, NM 87801-0387, USA}
\altaffiltext{6}{National Astronomical Observatory of Japan, Mitaka, Tokyo 181-8588, Japan}
\maketitle
\begin{abstract}
We have observed the compact H$_{\mathrm{II}}$ region complex nearest  to the dynamical center of the Galaxy, G-0.02-0.07,  using ALMA  in the H42$\alpha$ recombination line,  CS $J=2-1$, H$^{13}$CO$^+ J=1-0$, and SiO $v=0, ~J=2-1$ emission lines, and 86 GHz continuum emission. 
 The H$_{\mathrm{II}}$ regions HII-A to HII-C in the cluster are clearly resolved into a shell-like feature with a bright-half and a dark-half in the recombination line and continuum emission.  
 The absorption features in the molecular emission lines  show that HII-A, B and C are located on the near side of the 50 km s$^{-1}$ Molecular Cloud (50MC) but HII-D is located on the far side.
The electron temperatures and densities range $T_{\mathrm{e}}=5150-5920$ K and $n_{\mathrm{e}}=950-2340$ cm$^{-3}$, respectively.  The electron temperatures on the bright-half are slightly lower than those on the dark-half, while the electron densities on the bright-half are slightly higher than those on the  dark-half. 
 The H$_{\mathrm{II}}$ regions are located on the molecular filaments in the 50MC. They have already broken through the filaments and are growing in the surrounding molecular gas.  There are some shocked molecular gas components around the H$_{\mathrm{II}}$ regions.
 From line width of the H42$\alpha$ recombination line, the expansion velocities from HII-A to HII-D are  estimated to be $V_{\mathrm{exp}}=16.7$, $11.6$, $11.1$, and $12.1$ km s$^{-1}$, respectively.  
The expansion timescales from HII-A to HII-D are estimated to be  $t_\mathrm{age}\simeq1.4\times10^4, 1.7\times10^4$, $2.0\times10^4$,  and $0.7\times10^4$ years, respectively. 
The spectral types of the central stars from HII-A to HII-D are estimated to be O8V, O9.5V, O9V, and B0V, respectively. 
The positional relation among the H$_{\mathrm{II}}$ regions, the SiO molecule enhancement area, and Class-I maser spots  suggest that the shock wave caused by a cloud-cloud collision propagated along the line from HII-C to HII-A in the 50MC.  The shock wave would trigger the massive star formation.\end{abstract}

\section{Introduction}
Young and highly luminous clusters including the Arches cluster and the Quintuplet cluster have been found in the  Sagittarius A (Sgr A) region by IR observations  in the last three decades (e.g., \cite{Figer1999}).  
Although they should be birth in the parent  molecular clouds of the region, it is an open question what mechanism is responsible for the formation of such young massive clusters in the dense, warm, and turbulent molecular clouds of the Sgr A region  (e.g. \cite {Bally1987}, \cite{Oka1998}, \cite{Tsuboi1999}). 
It is difficult to demonstrate observationally how the molecular clouds produce such massive clusters because almost these clusters have already lost the surrounding molecular materials. 

The G-0.02-0.07 H$_{\mathrm{II}}$ region complex is a group of three compact and one ultra-compact H$_{\mathrm{II}}$ regions  in the Sgr A region.  This was first identified using Very Large Array (VLA) \citep{Ekers1983}.
The  complex is located only 3$'$ east in projection from the dynamical center of the Galaxy, Sagittarius A$^{\ast}$(Sgr A$^{\ast}$),   where is in the east part of the ``Galactic Center 50 km s$^{-1}$ Molecular Cloud" (50MC). 
In addition, because the position corresponds to the east limb of the nearest supernova remnant, Sagittarius A East (Sgr A-E), the 50 MC is believed to interact with it (e.g. \cite{Ho1985}, \cite{Tsuboi et al. 2009}).  
The H$_{\mathrm{II}}$ regions in the complex have radial velocities ranging of $43-49$ km s$^{-1}$ (\cite{Goss1985}, \cite{Serabyn1992}, \cite{Yusef-Zadeh2010}). The similar velocities suggest  that the complex is physically associated with the 50MC. The 50 MC has still abundant molecular gas (e.g.  \cite {Bally1987}, \cite{Oka1998}, \cite{Tsuboi1999}).  If the 50MC is the parent  molecular cloud in which the complex is recently formed, the structure and kinematics of the ionized gas of the G-0.02-0.07 H$_{\mathrm{II}}$ region complex and the surrounding molecular gas may trace the history of the  massive star formation. However, there has been no study to compare between these in arcsecond scale.  Atacama Large Millimeter/sub-millimeter Array (ALMA) can obtain continuum, recombination line, and molecular line data simultaneously.

We present new observational results with $\sim2\arcsec$ resolution using ALMA to solve this issue. Throughout this paper, we adopt 8 kpc as the distance to the Galactic center (e.g. \cite{Boehle}). Then, $1\arcsec$ corresponds to about 0.04 pc at the distance.  In addition, we use the Galactic coordinates. 

\section{Observation and Data Reduction}
We have performed the observation of the  G-0.02-0.07 H$_{\mathrm{II}}$ region complex in the  H42$\alpha$ recombination line ($85.6884$ GHz) and several molecular emission lines including  the H$^{13}$CO$^+ J=1-0$ ($86.754288$ GHz), SiO $v=0~J=2-1$ ($86.846995$ GHz), and  CS $J=2-1$ ($97.980953$ GHz) emission lines as a part of  ALMA Cy.1 observation (2012.1.00080.S. PI M.Tsuboi). 
The H42$\alpha$ recombination line is an ionized gas  tracer.  
  The CS emission line is a tracer of medium dense molecular gas, $n({\mathrm H}_2)_{\mathrm{cl}}\sim10^4$ cm$^{-3}$,   while the H$^{13}$CO$^+$ emission line is a tracer of dense molecular gas, $n({\mathrm H}_2)_{\mathrm{cl}}\sim10^5$ cm$^{-3}$. 
The SiO emission line is a tracer of strong C-shock  with $\Delta V\gtrsim 30$ km s$^{-1}$ in molecular clouds (e.g. \cite{May2000, Gusdorf, Jim}).
The entire ALMA observation consists of a 137 pointing mosaic of the 12-m array and a 52 pointing mosaic of the 7-m array (ACA), covering a $330\arcsec \times 330\arcsec$ area including  the G-0.02-0.07 H$_{\mathrm{II}}$ region complex and 50MC. 
 We have detected the G-0.02-0.07 H$_{\mathrm{II}}$ region complex in the H42$\alpha$ recombination line, the continuum emission at 86 GHz. We concentrate on the G-0.02-0.07 H$_{\mathrm{II}}$ region complex in this paper. 
 The angular resolutions using ``natural weighting" as {\it u-v} sampling at 86 and 98 GHz are $2.5\arcsec \times 1.9\arcsec, PA=-31^\circ$ and $2.2\arcsec \times 1.7\arcsec$, respectively.  While the angular resolutions using ``briggs weighting ($R=0.5$)" at 86 and 98 GHz are $2.0\arcsec \times 1.4\arcsec$ and $1.9\arcsec \times 1.3\arcsec$, respectively. 
The frequency channel width is 244 kHz. The velocity resolution is $1.7$ km s$^{-1}$(488 kHz).   J0006-0623, J1517-2422, J717-3342,  J1733-1304, J1743-3058, J1744-3116 and J2148+0657 were used as phase calibrators. The flux density scale was determined using Titan, Neptune and Mars. Because the observation has a large time span of one year and seven months,  the absolute flux density uncertainty  may be as large as 15 \%. The calibration and imaging of the data were done by CASA \citep{McMullin}. The continuum emission  was subtracted from the spectral data using the CASA task UVCONTSUB ($fitorder=1$).

We use the integrated intensity maps and channel maps in the molecular emission lines  to compare the distribution of the ionized gas with that of the molecular gas.  However, the molecular gas is significantly resolved out only by interferometer observations  when they have been processed in the same procedure mentioned above because the molecular gas is widely extended  in the area. The combining with the single-dish data, Total Power Array data, is required to recover the missing flux.  We  performed it  using the CASA task "FEATHER". 
We presented  the detailed description of the procedure and the full results in another paper \citep{Uehara}.  
\begin{figure}
\begin{center}
\includegraphics[width=17cm, bb=0 0 531.36 366.86]{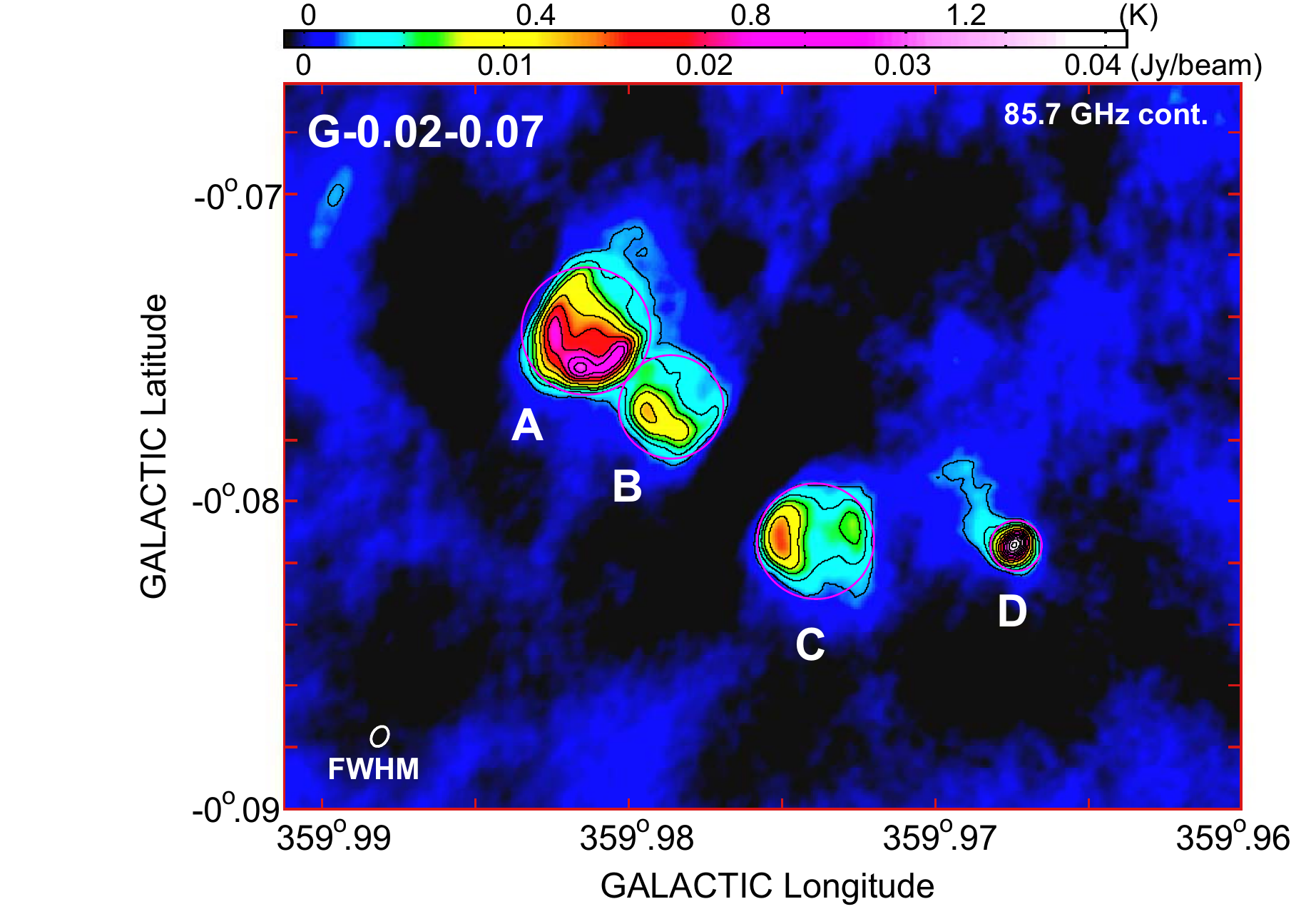}
\end{center}
\caption{Continuum map  of the compact H$_{\mathrm{II}}$ region complex G-0.02-0.07 in the Galactic Center with ALMA at 85.7 GHz. The H$_{\mathrm{II}}$ regions are also known as the Sagittarius A East H$_{\mathrm{II}}$ regions HII-A to HII-D (label A-D). The FWHM beam size is $\theta_{\mathrm{a}}\times\theta_{\mathrm{b}}=2.5\arcsec \times 1.9\arcsec$, ($PA=-31^\circ$) using ``natural weighting" and shown as an open oval in the bottom left corner. The contour levels are  $2, 4, 6, 8, 12,16, 20, 24,$ and $28$ mJy beam$^{-1}$ or $0.07,0.14, 0.21,0.28, 0.43, 0.57,0.71,0.85,$ and $0.99$ K in $T_{\mathrm B}$. The r.m.s noise is $0.5$ mJy beam$^{-1}$ or $0.018$ K in $T_{\mathrm B}$.  The red circles are the integration circles with the mean radii.
}
\end{figure}
\begin{figure}
\begin{center}
\includegraphics[width=8cm, bb=0 0 372.47 441.2 ]{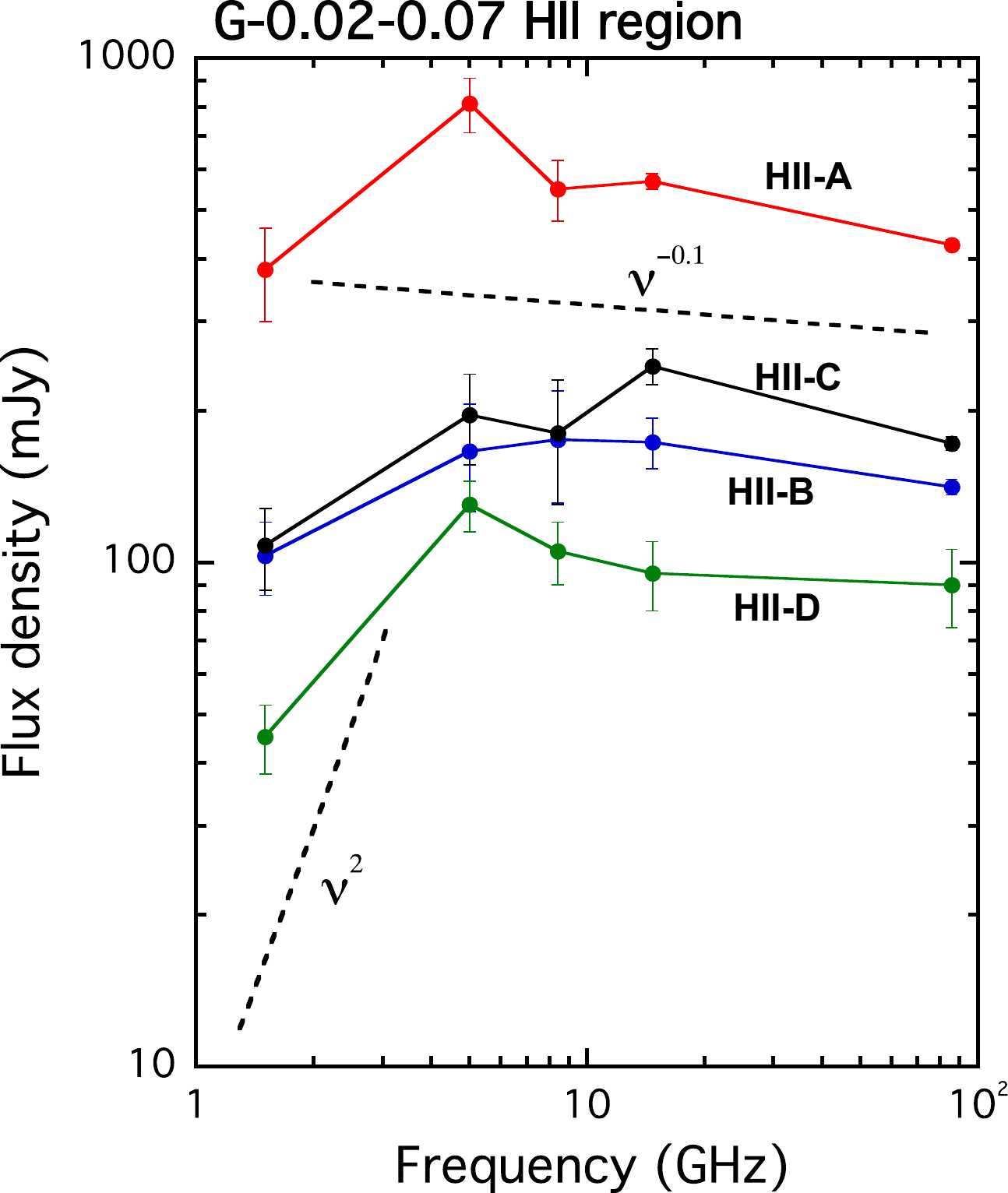}
\end{center}
\caption{Continuum spectra  of the compact H$_{\mathrm{II}}$ regions in G-0.02-0.07. The flux densities at 1.5 and 5 GHz are by Ekers et al. (1984). The flux densities at 8.4 and 14.7 GHz are by Mills et al. (2011) and Goss et al. (1985), respectively. The integration areas at 85.7 GHz is shown as red circles in Figure 1.}
\end{figure}
\section{Results}   
\subsection{Continuum Maps and Continuum Spectra of the G-0.02-0.07 Complex}
Figure 1 shows the continuum map at 85.7 GHz of the compact H$_{\mathrm{II}}$ region complex, G-0.02-0.07, in the Galactic Center with ALMA. The FWHM beam size is $\theta_{\mathrm{a}}\times\theta_{\mathrm{b}}=2.5\arcsec \times 1.9\arcsec$, ($PA=-31^\circ$) using ``natural weighting". 
The H$_{\mathrm{II}}$ regions are detected as compact objects distributed along the northeast-southwest direction. They are also known as the Sagittarius A East H$_{\mathrm{II}}$ regions A to D (hereafter HII-A to HII-D). 
HII-A, HII-B, and HII-C are clearly resolved into shell-like structures, which have a bright-half and dark-half. The structures have been reported in previous observations (e.g. \cite{Yusef-Zadeh1987}, \cite{Mills}).  The mean radius is defined by 
$\bar{r}{\mathrm{[pc]}}=0.04\times\sqrt{\mathrm{area}/\pi-\theta_{\mathrm{a}}\times\theta_{\mathrm{b}}/4}$. 
The area is within the 5 $\sigma$ contour. The mean radii of HII-A, HII-B, HII-C and  HII-D are $\bar{r}=0.30$, $0.24$,  $0.27$, and $0.12$ pc, respectively. 
Although HII-D has been reported to be resolved  into two continuum sources which are separated by $1.2\arcsec$ (\cite{Yusef-Zadeh2010}), These are not resolved probably because of the shortage of angular resolution of this observation. 

 We derived integrated intensities, $S_\nu$, and continuum brightness temperatures, $T_{\mathrm B}=1.22\times10^6\Big(\frac{\mathrm{area}/\pi-\theta_{\mathrm{a}}\times\theta_{\mathrm{b}}/4}{\mathrm{arcsec}^2} \Big)^{-1}\Big(\frac{\nu}{\mathrm{GHz}}\Big)^{-2}S_\nu$, at 85.7 GHz of the compact H$_{\mathrm{II}}$ regions.
The integration areas are shown as red circles in Figure 1. 
The flux densities and those at lower frequencies in previous observations (\cite{Ekers1983},  \cite{Goss1985}, \cite{Mills}) are summarized in Table 1. 
Figure 2 shows the continuum spectra  of the compact H$_{\mathrm{II}}$ regions. Although there are some scatters in these flux densities,  our derived values are on the whole consistent with  previous values at lower frequencies assuming that the continuum emission is mainly originated from the ionized gas through thin bremsstrahlung emission mechanism or $S_\nu\propto\nu^{-0.1}$.

\begin{figure}
\begin{center}
\includegraphics[width=15cm, bb=0 0 1009.5 1327.74]{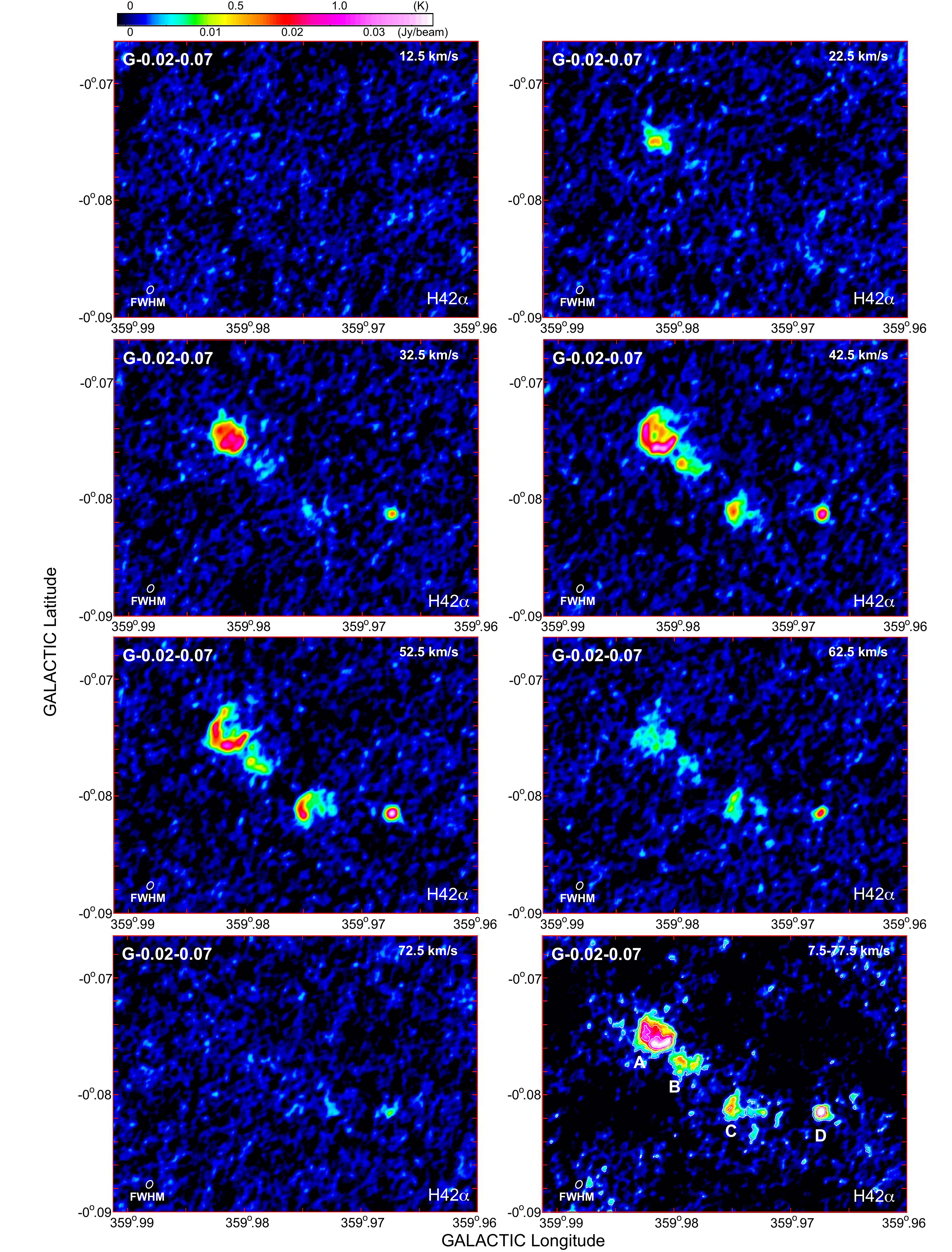}
\end{center}
\caption{Channel maps of the compact H$_{\mathrm{II}}$ region complex G-0.02-0.07  in the H42$\alpha$ recombination line. The angular resolution is $1.9\arcsec \times 1.3\arcsec$($PA=-37^\circ$) using ``Briggs weighting", which is shown on the lower-left corner as an open oval.   The central velocity is shown in the upper right corner. The velocity width is $\Delta V=10$ km s$^{-1}$. The r.m.s noise is $1.3$ mJy beam$^{-1}$ or $0.05$ K in $T_{\mathrm B}$.
 The bottom right panel shows the integrated intensity map  of the H42$\alpha$ recombination line. The integrated velocity range is from $V_{\mathrm{LSR}}=7.5$ to $77.5$ km s$^{-1}$. The contour levels are  $0.2, 0.4, 0.6, 0.8,$ and $1.0$ Jy beam$^{-1}$ km s$^{-1}$ or $7, 14, 21, 28$ and $35$ K km s$^{-1}$. The r.m.s noise is $0.1$ Jy beam$^{-1}$  km s$^{-1}$ or $3.5$ K km s$^{-1}$.  }
\end{figure}

\subsection{Channel Maps in the H42$\alpha$ recombination line}
Figure 3 shows channel maps of the compact H$_{\mathrm{II}}$ region complex G-0.02-0.07  in the H42$\alpha$ recombination line. The angular resolution is $2.5\arcsec \times 1.9\arcsec$($PA=-31^\circ$) using ``natural weighting".  The component of HII-A is small at 
the smallest velocity of 22.5 km s$^{-1}$ 
and becomes larger and shapes a half shell-like feature with increasing velocity.
The component becomes small and faint at the positive end velocity.  Similar tendencies are also seen in HII-B and HII-C.  Therefore, we consider that these H$_{\mathrm{II}}$ regions have half-shell-like structures in the $l-b-v$ space. On the other hand, HII-D shows compact peaks in all the maps from $V_{\mathrm{LSR}}=32.5$ to $62.5$ km s$^{-1}$.

The bottom right panel of Figure 3 shows the integrated intensity map  of the H42$\alpha$ recombination line. The  integrated velocity range is from $V_{\mathrm{LSR}}=7.5$ to $77.5$ km s$^{-1}$.  The total intensity map is similar to the continuum one of Figure 1. The integrated line intensities of HII-A, HII-B, HII-C, and HII-D are $\int S_{\mathrm{line}}(\mathrm{H}42\alpha)dV=15.8\pm0.5$, $6.1\pm0.2$, $6.2\pm0.2$, and $3.6\pm0.3$ Jy km s$^{-1}$, respectively. The integration areas are shown as red circles in Figure 1. The  integrated line intensities are also summarized in Table 1.
In the continuum map at 85.7 GHz(see Figure 1), HII-D has a faint component extending to northeast.  However, the component is not identified in the integrated intensity map of the H42$\alpha$ recombination line (see Figure 3). Then this component in the continuum map would be an artifact by the contamination of molecular emission lines.

\begin{figure}
\begin{center}
\includegraphics[width=15cm, bb=0 0 1000 1342.01]{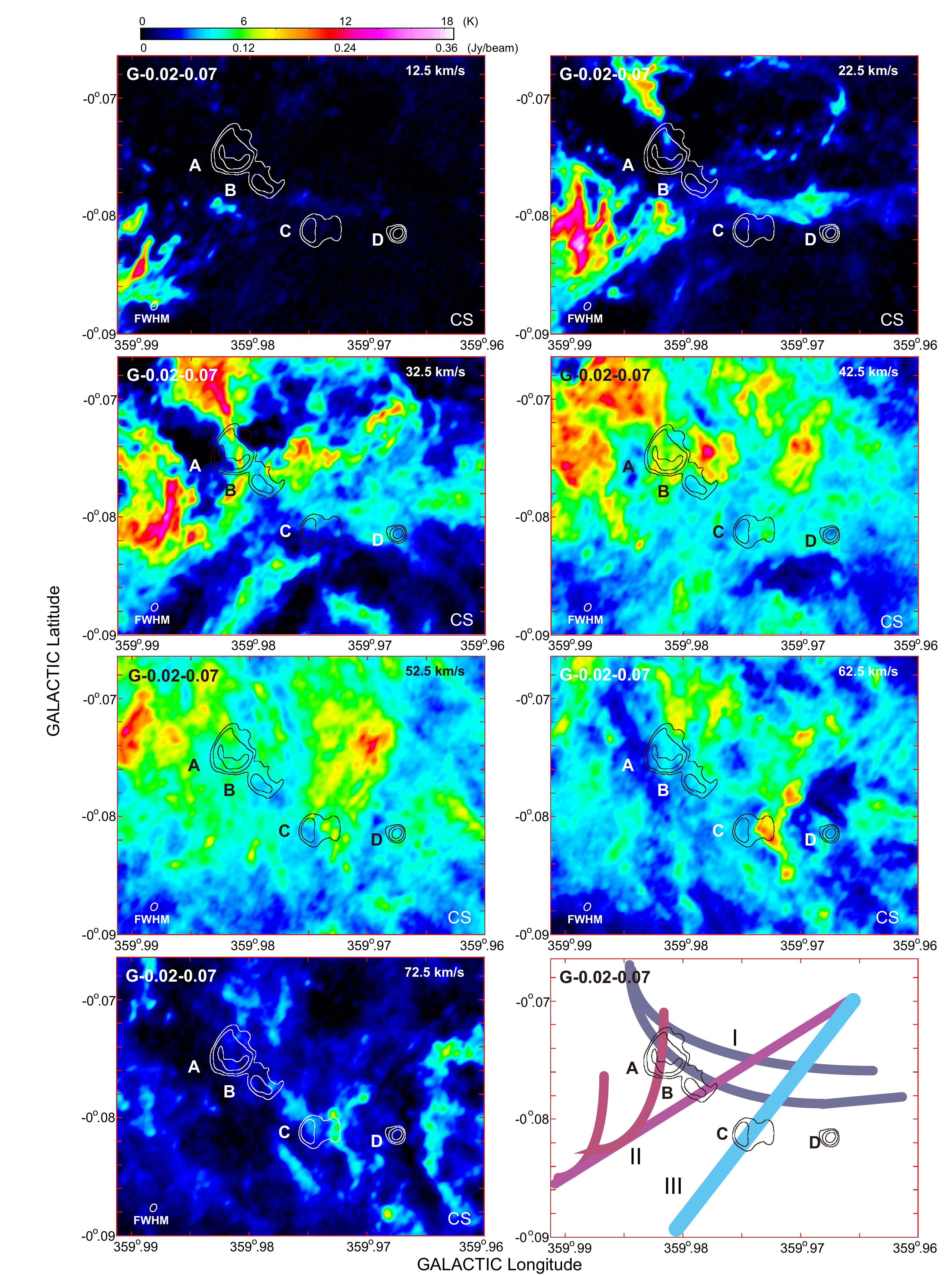}
\end{center}
\caption{Channel maps of the compact H$_{\mathrm{II}}$ region complex G-0.02-0.07  in the CS $J=2-1$ emission line. The angular resolution is $1.9\arcsec \times 1.3\arcsec$($PA=-37^\circ$) using ``Briggs weighting", which is shown on the lower-left corner as an open oval.   The central velocity is shown in the upper right corner. The velocity width is $\Delta V=10$ km s$^{-1}$. The r.m.s noise is $1.9$ mJy beam$^{-1}$ or $0.10$ K in $T_{\mathrm B}$. Contours show the continuum map at 85.7 GHz shown in Figure 1 for comparison. The bottom right panel shows the finding chart of molecular filaments. }
\end{figure}
\begin{figure}
\begin{center}
\includegraphics[width=15cm, bb=0 0 1015.5 1356.15]{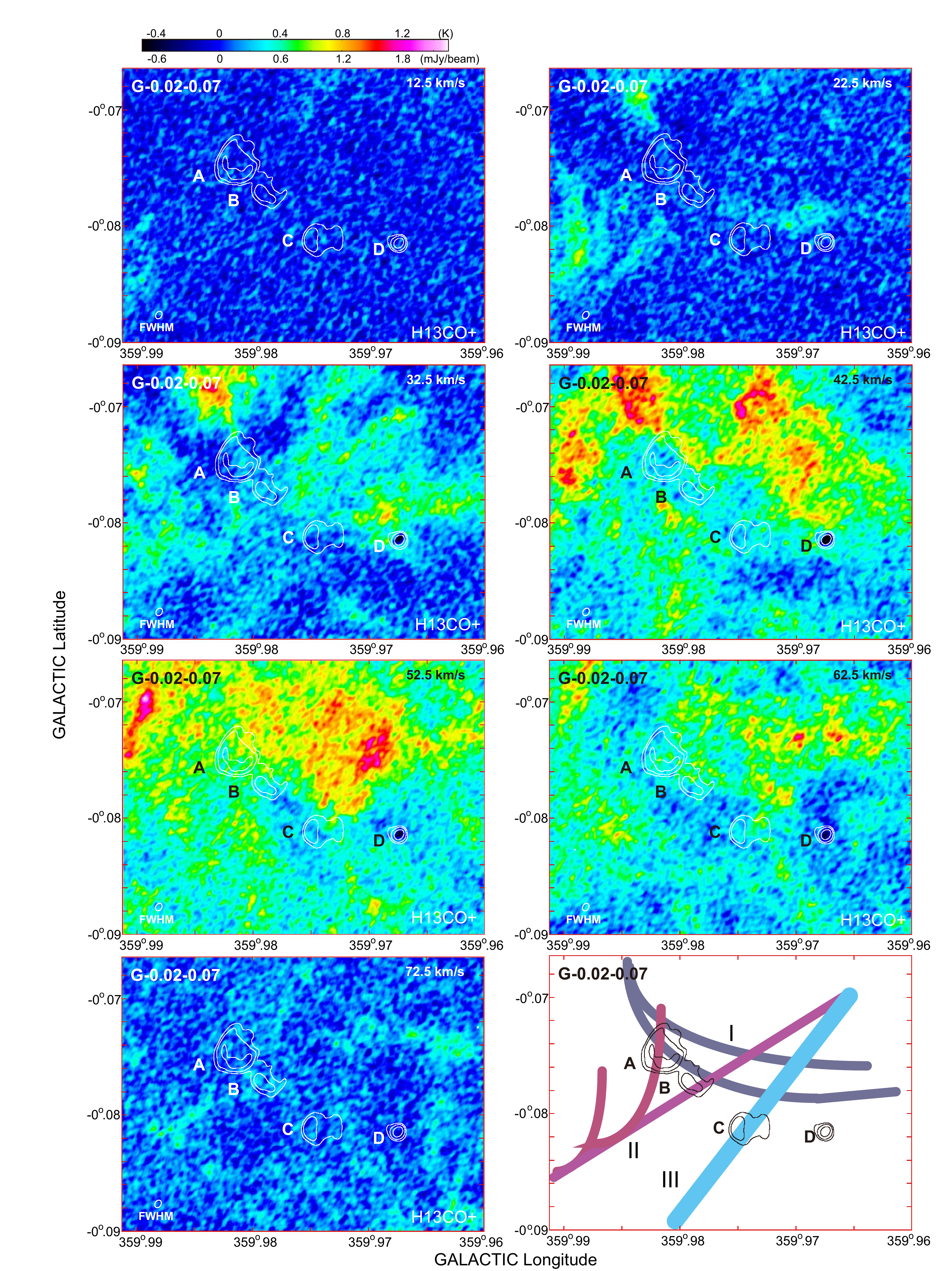}
\end{center}
\caption{Channel maps of the compact H$_{\mathrm{II}}$ region complex G-0.02-0.07  in the H$^{13}$CO$^+ J=1-0$. 
The angular resolution is $2.0\arcsec \times 1.4\arcsec$($PA=-39^\circ$) using ``Briggs weighting", which is shown on the lower-left corner as an open oval.   The central velocity is shown in the upper right corner. 
The velocity width is $\Delta V=10$ km s$^{-1}$. The r.m.s noise is $1.4$ mJy beam$^{-1}$ or $0.08$ K in $T_{\mathrm B}$. Contours show the continuum map at 85.7 GHz shown in Figure 1 for comparison. The bottom right panel shows the finding chart of molecular filaments. 
  }
\end{figure}
\begin{figure}
\begin{center}
\includegraphics[width=15cm, bb=0 0 1001.71 1342.01]{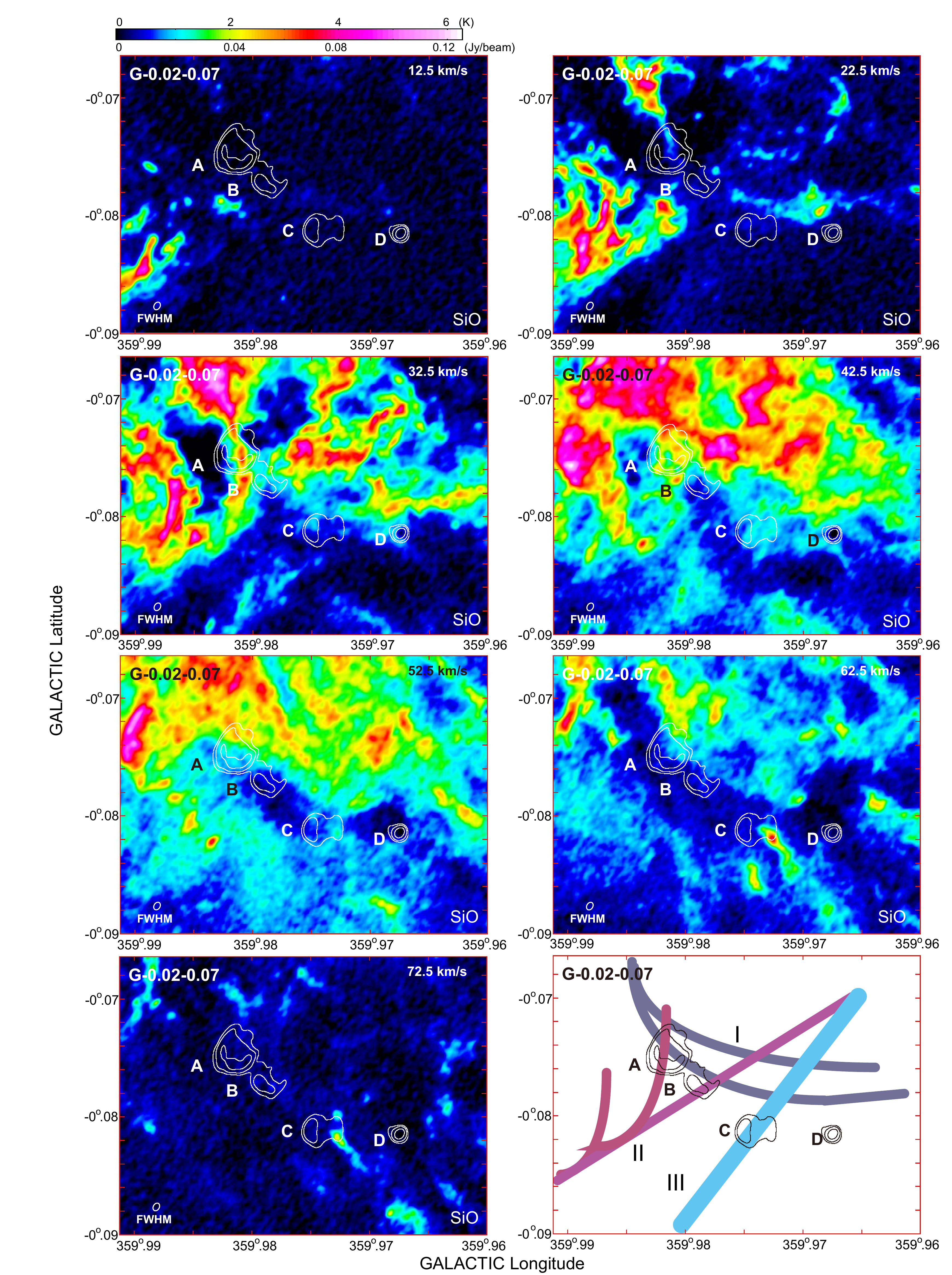}
\end{center}
\caption{Channel maps of the compact H$_{\mathrm{II}}$ region complex G-0.02-0.07  in the SiO $v=0, J=2-1$. 
The angular resolution is $2.0\arcsec \times 1.4\arcsec$($PA=-39^\circ$) using ``Briggs weighting", which is shown on the lower-left corner as an open oval.   The central velocity is shown in the upper right corner. 
The velocity width is $\Delta V=10$ km s$^{-1}$. The r.m.s noise is $1.4$ mJy beam$^{-1}$ or $0.08$ K in $T_{\mathrm B}$. Contours show the continuum map at 85.7 GHz shown in Figure 1 for comparison. The bottom right panel shows the finding chart of molecular filaments. 
  }
\end{figure}

\subsection{Channel Maps in the Molecular Emission Lines}
Figure 4, Figure 5, and Figure 6 show channel maps of the compact H$_{\mathrm{II}}$ region complex G-0.02-0.07 in the CS $J=2-1$, H$^{13}$CO$^+ J=1-0$, and SiO $v=0, J=2-1$ emission lines, respectively. The contours show the continuum map at 85.7 GHz shown in Figure 1 for comparison. The angular resolutions using ``Briggs weighting are $1.9\arcsec \times 1.3\arcsec$($PA=-37^\circ$), $2.0\arcsec \times 1.4\arcsec$($PA=-39^\circ$), and $2.0\arcsec \times 1.4\arcsec$($PA=-39^\circ$), respectively. The molecular filaments entangled with several blobs are identified in these maps.
A curved molecular filament is identified to be extend from $l\sim359^\circ.983, b\sim-0^\circ067$ to $l\sim359^\circ.965, b\sim-0^\circ080$ in the CS and SiO maps from $V_{\mathrm{LSR}}=22.5$ to $42.5$ km s$^{-1}$ (hereafter Filament I).  Another  molecular filament seems to  extend from $l\sim359^\circ.990, b\sim-0^\circ084$ to $l\sim359^\circ.965, b\sim-0^\circ070$ in the CS maps from $V_{\mathrm{LSR}}=22.5$ to $42.5$ km s$^{-1}$ (hereafter Filament II).  In the CS maps from $V_{\mathrm{LSR}}=32.5$ to $62.5$ km s$^{-1}$, a broad  molecular filament seems to  extend from $l\sim359^\circ.982, b\sim-0^\circ090$ to $l\sim359^\circ.968, b\sim-0^\circ070$ (hereafter Filament III). The bottom right panel shows the finding chart of these molecular filaments. 
In addition, several minor molecular filaments may be seen in all the CS maps from $V_{\mathrm{LSR}}=12.5$ to $72.5$ km s$^{-1}$.  Although the appearances of the molecular gas components in the SiO  emission line roughly resemble those in the CS emission line, the filaments mentioned above seem to be emphasized in the SiO emission line. While these molecular filaments are not clear in the H$^{13}$CO$^+$ maps although filament I can be identified.

HII-A is located on the crossing part of the filaments I and II.  The part corresponding to HII-A is dimmed comparing with the other parts of the filament I in the CS maps from $V_{\mathrm{LSR}}=22.5$ to $42.5$ km s$^{-1}$.  While the HII-A part of the filament II is brighten rather than dimmed in the SiO map of $V_{\mathrm{LSR}}=32.5$ km s$^{-1}$. Such correlation and anti-correlation are not clear in the H$^{13}$CO$^+$ maps.
HII-B is located on another crossing part of the filaments I and II.  This part is also dimmed comparing with the other parts of the filaments in the CS maps from $V_{\mathrm{LSR}}=22.5$ to $42.5$ km s$^{-1}$.  The dimming is also seen in the SiO maps. 
HII-C seems to be located in a faint part of the the filament III in the CS maps from $V_{\mathrm{LSR}}=42.5$ to $62.5$ km s$^{-1}$. 
These morphological compensations between the molecular gas and the ionized gas are caused by physical association or by absorption of the continuum emission. This issue will be examined in Discussion.

A deep absorption feature corresponding to HII-D is seen in the maps in the H$^{13}$CO$^+ J=1-0$ emission line from  $V_{\mathrm{LSR}}=32.5$  to $62.5$  km s$^{-1}$.   The absorption feature of HII-D is also identified in the CS and SiO maps of $42.5$ km s$^{-1}$ although these are  shallower than that in the H$^{13}$CO$^+ J=1-0$ maps (also see \cite{Uehara}). While HII-D does not seem to be associated with any molecular filaments.

\subsection{Line Profiles of the G-0.02-0.07 Complex} 
Figure 7a shows line profiles of the H42$\alpha$ recombination line toward HII-A, HII-B, HII-C, and HII-D. The integration areas are shown as the red circles in Figure 1.  The compact H$_{\mathrm{II}}$ regions are identified as single peaked profiles. The LSR center velocities  by Gaussian fit of HII-A, HII-B, HII-C, and HII-D are $V_{\mathrm{LSR, C}}=42.7\pm0.3$, $46.2\pm0.7$, $49.4\pm0.3$, and $49.2\pm0.8$ km s$^{-1}$, respectively. Meanwhile the FWHM velocity widths of HII-A, HII-B, HII-C, and HII-D are $\Delta V_{\mathrm{FWHM}}=36.2\pm1.6$, $26.6\pm2.6$, $26.2\pm1.2$, and $27.6\pm2.9$ km s$^{-1}$, respectively. These derived velocities are consistent with those of previous observations (e.g. \cite{Goss1985}, \cite{Mills}). The velocities are also summarized in Table 1. 
The counterparts of the He42$\alpha$ recombination line should be observed at $V_{\mathrm{LSR, C}}\sim-80$ km s$^{-1}$ or shifted by $\Delta V\sim-120$ km s$^{-1}$ from the peak in the H42$\alpha$ recombination line (arrows in Figure 4a). Those of HII-A, HII-B, and HII-C are detected although they are faint.  The recombination line intensity ratio of  He to H is $\frac{T_\mathrm{B}(\mathrm{He})}{T_\mathrm{B}(\mathrm{H})}\sim0.1$, which are consistent with those observed usually in the Galactic disk region (e.g. \cite{Rubin}). While the counterpart of HII-D is not detected. This difference may be caused by the spectral type of the central star (see Discussion).

Figure 7b shows line profiles of the CS $J=2-1$ emission line  toward these compact H$_{\mathrm{II}}$ regions. The integration areas are the same as those of the H42$\alpha$ recombination line. The CS line profiles of HII-B, HII-C, and HII-D have  shallow dips around the peaks velocities of the recombination line although that of HII-A is slightly unclear. These dips may be signatures made by absorption of the background continuum emission.
While Figure 7c shows line profiles of the H$^{13}$CO$^+ J=1-0$ emission line  toward these compact H$_{\mathrm{II}}$ regions. The integration areas are the same as those of the H42$\alpha$ recombination line. 
HII-A has a Gaussian-like single peaked profile in the H$^{13}$CO$^+ $ emission line.  
HII-B and HII-C have broad peaks with some irregularities around the peak velocities of the recombination line.
The line profile of HII-D is deformed significantly.  In addition, the broken line in Figure 7c also shows the line profile within the central $3\arcsec$ of HII-D. 
This profile has a deep dip around the peak velocity of the recombination line, which would be made by absorption of the background continuum emission toward HII-D. These features in the molecular emission lines will be discussed in the next section.

Figure 7d shows line profiles of the SiO $J=2-1$ emission line  toward these compact H$_{\mathrm{II}}$ regions.
The peak velocities of the SiO $J=2-1$ emission line seem to correspond to the smaller halves of the velocity ranges of the other emission lines.
These would mean that there are SiO enhanced parts in the smaller velocity side of the cloud. 

\begin{figure}
\begin{center}
\includegraphics[width=17.5cm, bb=0 0 1265.81 610.28]{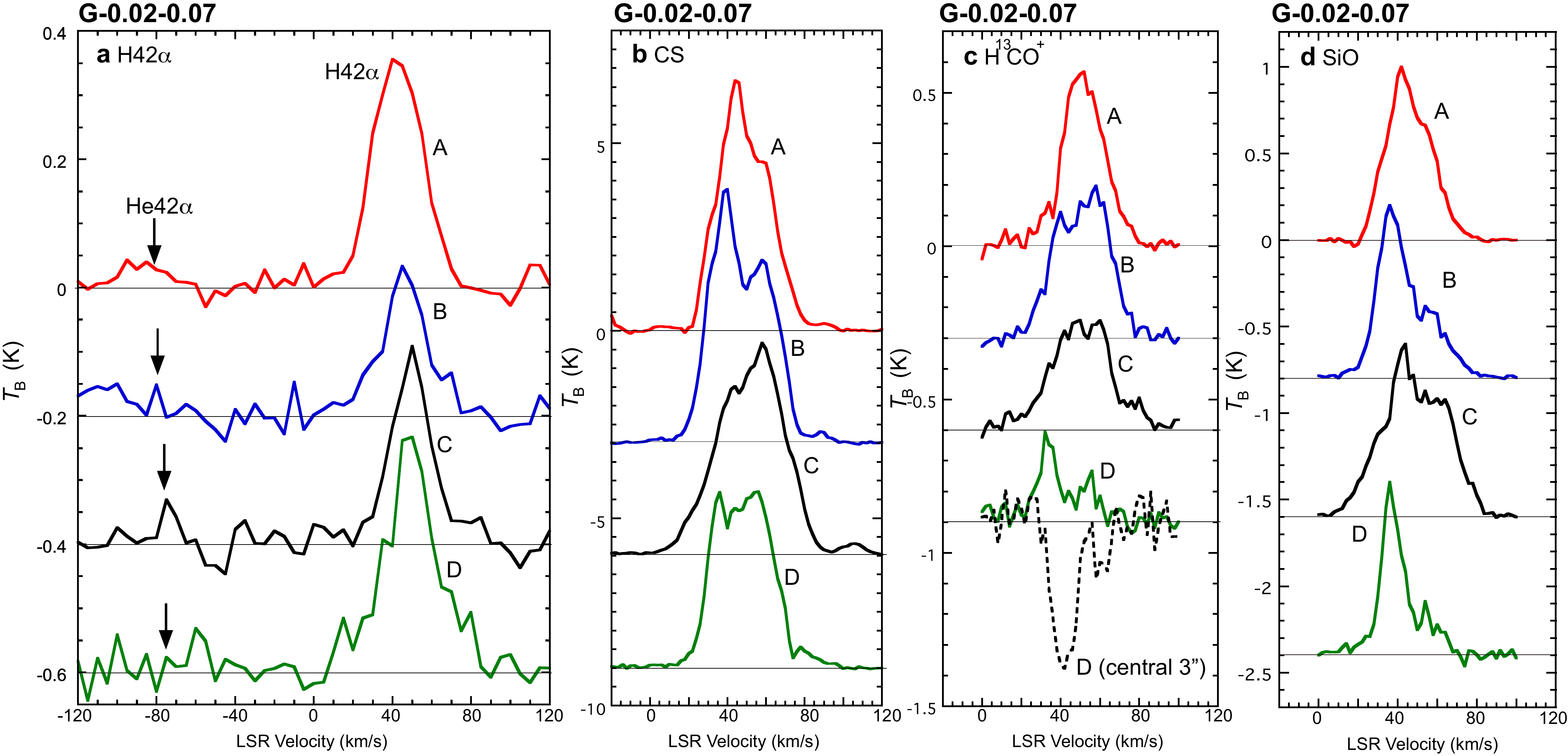}
\end{center}
\caption{{\bf a} Line profiles of the H42$\alpha$ recombination line (thick lines)  toward the compact H$_{\mathrm{II}}$ regions in G-0.02-0.07. The integration areas of both lines are shown as red circles in Figure 1. The arrows show the velocities of the counterparts of the He42$\alpha$ recombination line. {\bf b} Line profiles of the CS $J=2-1$ emission line toward the same areas. {\bf c} Line profiles of the  H$^{13}$CO$^+ J=1-0$ emission line toward the same areas. In addition, the line profile within the central $3\arcsec$ of HII-D is also shown (broken line). {\bf d} Line profiles of the SiO $v=0, J=2-1$ emission line toward the same areas.}
\end{figure}

\subsection{Position-velocity Diagrams of the H42$\alpha$ recombination line across the Compact HII regions}
Figure 8 shows position-velocity (PV) diagrams  of the H42$\alpha$ recombination line across the compact H$_{\mathrm{II}}$ regions of  G-0.02-0.07. The sampling areas are shown as red rectangles in the guide maps of the panels.  Figure 8a is the PV diagram crossing HII-A and HII-B.  Figure 1 shows that HII-A and HII-B have similar  continuum asymmetric distributions; a half of limb-brightening and the other dimming half.  The position-offset axis of the PV diagram is perpendicular to the asymmetric distribution.  The feature corresponding to HII-A is identified as a U-shaped feature in the diagram. The positive offset side of this feature is brighter than the negative offset side. The velocity extent of HII-A is  $\Delta V_{\mathrm{FWZI}}\simeq 50$ km s$^{-1}$. On the other hand, the feature corresponding to HII-B is clear on the negative offset side but dimming on the positive offset side.

Figures 8b and 8c are also the PV diagrams of HII-A and HII-B, respectively. The offset axes of the diagram at $PA=-40^{\circ}$ are parallel to the asymmetric distributions of HII-A and HII-B.  A U-shaped feature is also identified in HII-A.  
The negative offset sides of HII-A and HII-B are brighter than the positive offset sides. 
These facts seem to be consistent with our interpretation that these H$_{\mathrm{II}}$ regions have half-shell-like structures in the $l-b-v$ space as mentioned in the previous subsection. 
The velocity extents of HII-A and HII-B in these diagrams are  $\Delta V_{\mathrm{FWZI}}\simeq 50$ and $45$ km s$^{-1}$, respectively.  Similar features have been reported in the PV diagram of the [Ne$_{\mathrm{II}}$] emission line (Fig.8 in \cite{Yusef-Zadeh2010}).

Figure 8d is the PV diagram crossing HII-C and HII-D. The offset axis of the diagram at $PA=90^{\circ}$ is parallel to the axis of the asymmetric distribution of HII-C. 
The negative offset side of HII-C is brighter than the positive offset side. 
HII-D is not spatially resolved in the diagram because of the shortage of the angular resolution.
The velocity extents of HII-C and HII-D are  $\Delta V_{\mathrm{FWZI}}\simeq 40$ km s$^{-1}$ and $50$ km s$^{-1}$, respectively. 
These velocities are also summarized in Table 1.

\begin{figure}
\begin{center}
\includegraphics[width=16cm, bb=0 0 855.1 808.97]{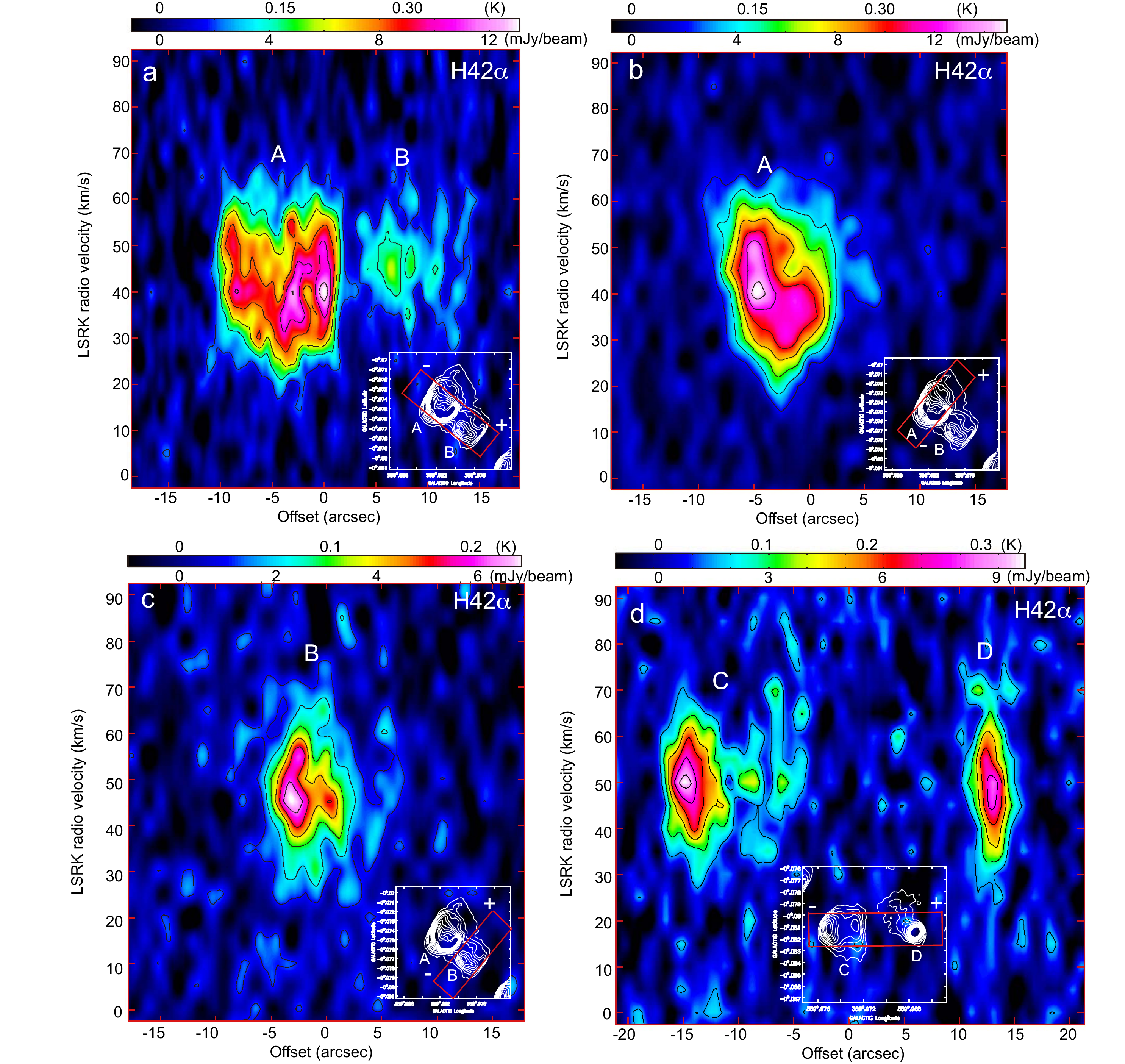}
\end{center}
\caption{Position-velocity diagrams the compact H$_{\mathrm{II}}$ regions in G-0.02-0.07 in the H42$\alpha$ recombination line. {\bf a} Position-velocity diagram along the row of HII-A and HII-B. The sampling area is shown as a red rectangle in the guide map  (contours). {\bf b} Position-velocity diagram of HII-A. The sampling area is shown as a red rectangle in the guide map  (contours). {\bf c} Position-velocity diagram of HII-B. The sampling area is shown as a red rectangle in the guide map  (contours).{\bf d} Position-velocity diagram along the row of HII-C and HII-D. The sampling area is shown as a red rectangle in the guide map  (contours). The "+" and "-" in the guide map show the direction of angular offset.}
\end{figure}

\begin{table}
  \caption{Physical Parameters of  the Compact H$_{\mathrm{II}}$  Regions in G-0.02-0.07.  }
  \label{tab:first}
 \begin{center}
    \begin{tabular}{ccccc}
 \hline   \hline
Region & HII-A &HII-B&HII-C& HII-D\\
\hline
$S_\nu$(1.5 GHz) [mJy]$^1$& $380\pm80$  &$103\pm17$&$108\pm20$& $45\pm7$ \\
$S_\nu$(5.0 GHz) [mJy]$^1$& $812\pm108$&$166\pm40$&$196\pm40$& $130\pm15$ \\
$S_\nu$(8.4 GHz) [mJy]$^2$&$550\pm75$&$175\pm44$&$180\pm50$& $105\pm15$ \\
$S_\nu$(14.7 GHz) [mJy]$^3$&$570\pm20$&$173\pm20$&$245\pm20$& $95\pm15$ \\
$S_\nu$(85.7 GHz) [mJy]&$426\pm10$&$141\pm5$&$172\pm5$& $90\pm10$ \\
\hline
$T_{\mathrm B}$(85.7 GHz) [K]&$0.316\pm0.007$&$0.157\pm0.006$&$0.152\pm0.004$& $0.415\pm0.046$ \\
\hline
Mean raidus $\bar{r}$ [pc]&$0.30$&$0.24$&$0.27$&$0.12$\\
\hline
$V_{\mathrm C}(\mathrm{H}42\alpha)$ [km s$^{-1}$]& $42.7\pm0.3$ &$46.2\pm0.7$&$49.4\pm0.3$& $49.2\pm0.8$ \\
$\Delta V_{\mathrm{FWHM}}(\mathrm{H}42\alpha)$ [km s$^{-1}$]$^4$& $36.2\pm1.6$& $26.6\pm2.6$&$26.2\pm1.2$& $27.6\pm2.9$\\ 
$\Delta V_{\mathrm{FWZI}}(\mathrm{H}42\alpha)$ [km s$^{-1}$]$^5$& $50$& $45$&$40$& $50$\\ 

\hline
$\int S_{\mathrm{line}}(\mathrm{H}42\alpha)dV$ [Jy km s$^{-1}$]$^6$& $15.8\pm0.5$ &$6.1\pm0.2$ &$6.2\pm0.2$ & $3.6\pm0.3$\\
\hline
Electron temperature $\bar{T}^\ast_{\mathrm{e}}$ [K]& $5780$ &$5150$ &$5920$ & $5470$\\
Electron density $\bar{n}_{\mathrm{e}}$ [cm$^{-3}$]& $1300$ &$1000$ &$950$ & $2340$\\
$\bar{EM}$ [pc cm$^{-6}$]& $6.8\times10^{5}$ &$3.3\times10^{5}$&$3.3\times10^{5}$ & $8.8\times10^{5}$\\
Ambient molecular gas density $\bar{n_0}$ [cm$^{-3}$]& $1.5\times10^4$ &$1.5\times10^4$&$1.5\times10^4$ & $1.5\times10^4$\\
\hline
Sound velocity $C_{\mathrm{s}}$[km$^{-1}$]$^7$& 9.9 &9.3&10.0 & 9.6\\
Expanding velocity $V_{\mathrm{exp}}$ [km$^{-1}$]$^8$&16.7&11.6&11.1&12.1\\
Str\"omgren radius $R_{\mathrm{S}}$ [pc]$^9$&$0.06$&$0.04$&$0.04$&$0.03$\\
Age $t_\mathrm{exp}$ [yr]$^{10}$& $1.4\times 10^4$ &$1.7\times 10^4$&$2.0\times 10^4$ & $0.7\times 10^4$\\
\hline
Ionized photon rate, $Q_0$ [s$^{-1}$]$^{11}$ &$2.3\times10^{48}$&$0.8\times10^{48}$&$0.9\times10^{48}$&$0.5\times10^{48}$\\
Spect. type & O8V &O9.5V &O9V & B0V\\
\hline
    \end{tabular}
 \end{center}
 $^1$ \cite{Ekers1983}.
$^2$ \cite{Mills}. $^3$  \cite{Goss1985}. $^4$ FWHM means full-width at half maximum. $^5$ FWZI means full-width at zero intensity. $^6$ The integrated velocity range is from $V_{\mathrm{LSR}}=7.5$ to $77.5$ km s$^{-1}$.  $^7~C_\mathrm{s}=(2kT^\ast_\mathrm{e}/m_\mathrm{H})^{0.5}$. $^8~V_{\mathrm{exp}}=\frac{1}{2}(\Delta V_{\mathrm{FWHM}}^2-(2\sqrt{\ln2}C_\mathrm{s})^2)^{0.5}$. $^9$ $R_\mathrm{s}\sim \Big(\frac{n_\mathrm{e}}{n_\mathrm{0}}\Big)^{2/3} \bar{r}$. $^{10}$ $t_\mathrm{exp}\sim\frac{\bar{r}-R_{\mathrm{S}}}{V_{\mathrm{exp}}}$. $^{11}$ $Q_0 = \frac{4\pi}{3}R_\mathrm{s}^3 \alpha_Bn_0^2$.
\clearpage
\end{table}

\section{Discussion}
\subsection{Positions of the H$_{\mathrm{II}}$ regions A-D on the Line-of-sight}
 Although it would be certain that the H$_{\mathrm{II}}$ regions A-D are associated physically with the 50MC, there is a controversy over where they are located on the line-of-sight referencing to the cloud.
One view is that the H$_{\mathrm{II}}$ regions A-C are located on the near side of the 50MC, which was claimed by
by Mills et al. being based on the extinction (2011). Another view is that they are located on the far side,  which is based on their apparent direction of the motions towards us \citep{Yusef-Zadeh2010}. 

As well known, the brightness temperature of a molecular emission line referencing to line-free frequency, $T_{\mathrm{B}}$, is given by
\begin{equation}
\label{ 1}
T_{\mathrm{B}}=\Big(T_{\mathrm{MC, ex}}-T_{\mathrm{CBR}}\Big)\Big(1-e^{-\tau_{\mathrm{MC}}}\Big)-T_{\mathrm{HII, cont}}\Big(1-e^{-\tau_{\mathrm{front}}}\Big)
\end{equation}
where $T_{\mathrm{MC, ex}}$, $T_{\mathrm{HII, cont}}$, $\tau_{\mathrm{MC}}$, and $\tau_{\mathrm{front}}$ are the excitation temperature of the molecular line, the continuum brightness temperature of  the H$_{\mathrm{II}}$ region,  the line optical thickness  of the whole molecular cloud, and the line optical thickness in front of  the H$_{\mathrm{II}}$ region, respectively. 

The $T_{\mathrm{HII, cont}}$ values toward the H$_{\mathrm{II}}$ regions A-C are lower than $\lesssim0.3$ K (see Figure 1 and Table 1), while the  $T_{\mathrm{B}}$ values of the CS emission line are higher than $\gtrsim5$ K (see Figure 4 and Figure 7b).  The term $\Big(T_{\mathrm{MC, ex}}-T_{\mathrm{CBR}}\Big)\Big(1-e^{-\tau_{\mathrm{MC}}}\Big)$ is much larger than $T_{\mathrm{HII, cont}}\Big(1-e^{-\tau_{\mathrm{front}}}\Big)$.  Thus a deep dip  in the CS line profile cannot be made by the absorption even in the case that the CS emission line is optically thick and/or the H$_{\mathrm{II}}$ region is located on the far side of the cloud. Therefore the remarkable dip at $V=50$ km s$^{-1}$ of HII-B in the CS emission line would be caused by kinematic components with $V=35$ and $55$ km s$^{-1}$ in the cloud (see Figure 10c) not by the absorption.

On the other hand, the  $T_{\mathrm{B}}$ values of the H$^{13}$CO$^+$ emission line are as high as 0.5 K (see Figure 5 and Figure 7c). Then the absorption of the continuum emission can make a deep dip when the emission line is optically thick and/or the H$_{\mathrm{II}}$ region is located on the far side of the cloud. The peaks toward HII-A and HII-C in the H$^{13}$CO$^+$ emission lines have no dips deeper than 0.02 K. 
While the dip at $V\sim50$ km s$^{-1}$ of HII-B in the H$^{13}$CO$^+$  emission line is seen. However this feature would be caused by kinematic structures of the cloud as mentioned above. 
These indicate that the H$^{13}$CO$^+$ emission line is optically thinner than $\tau_{\mathrm{front}} \lesssim0.1$ or the H$_{\mathrm{II}}$ regions A-C are located on the near side of the cloud. This is consistent with that they have no such large extinction in $A_{\mathrm{v}}$ (\cite{Mills}) and no corresponding deep absorption (\cite{Uehara}). 

The continuum emission, $T_{\mathrm{HII, B}}$, toward HII-D is $0.42\pm0.05$ K (see Figure 1 and Table 1).  The  $T_{\mathrm{B}}$ values of the CS and H$^{13}$CO$^+$ emission lines are 4.5 and 0.3 K, respectively (see Figure 7b and Figure 7c). A shallow dip is identified around the peak of the CS emission line. While this feature is also identified as a deep dip, $\Delta T \sim 0.3$ K,  in the H$^{13}$CO$^+$ emission line. Therefore the deep dip indicates that the molecular cloud is estimated to be moderate optically-thick, $\tau_{\mathrm{front}} \sim 1$, in the H$^{13}$CO$^+$ emission line. If HII-D is located on the near side of the 50MC, the thickness of the whole cloud, $\tau_{\mathrm{MC}}$, becomes thicker and the $T_{\mathrm{B}}$ becomes higher like  the CS emission line. Therefore HII-D is located on the far side of the 50MC. 
This is consistent with the larger extinction of $A_{\mathrm{v}}\sim71$ mag measured toward HII-D (\cite{Mills}).

\subsection{Electron Temperature and Electron Density of the H$_{\mathrm{II}}$ regions A-D}
\subsubsection{Electron Temperature}
The LTE electron temperature, $T^\ast_{\mathrm e}$, of the H$_{\mathrm{II}}$ regions A-D is estimated from the ratio between the integrated recombination line intensity, $\int S_{\mathrm{line}}(\mathrm{H}42\alpha)dV $, and the continuum flux density, $S_\nu(\mathrm{85.7GHz})$, assuming that the line and continuum emissions are optically thin. 
As mentioned above, the compact H$_{\mathrm{II}}$ regions have single-peak line profiles,  thus it is easy to derive the electron temperature at these positions.
The well-known formula of the LTE electron temperature is given by
\begin{equation}
\label{2}
T^\ast_{\mathrm e}[\mathrm K]=\left[\frac{6.985\times10^3}{a(\nu, T^\ast_{\mathrm e})}\Big(\frac{\nu}{\mathrm{GHz}}\Big)^{1.1}
\frac{1}{1+\frac{N(\mathrm{He^+})}{N(\mathrm{H^+})}}
\frac{S_\nu(\nu)}
{\int S_{\mathrm{line}}\Big(\frac{dV}{\mathrm{km~s}^{-1}}\Big)}
\right]^{\frac{1}{1.15}}.
\end{equation}
The correction factor, $a(\nu, T^\ast_{\mathrm e})$, for $\nu=85.7$ GHz and $T^\ast_{\mathrm e}=4\times10^3-1.5\times10^4$ K is $0.822-0.942$ \citep{Mezger}.  We assume that the number ratio of He$^+$ to H$^+$ is $\frac{N(\mathrm{He^+})}{N(\mathrm{H^+})}=0.09$, a typical value for the Orion A HII region (e.g. \cite{Rubin}). This is consistent with the faint detection of the He42$\alpha$ recombination line mentioned above (see Figure 7a).
The mean electron temperatures of HII-A, HII-B, HII-C, and HII-D are $\bar{T}^\ast_{\mathrm e}=5780$, $5150$, $5920$, and $5470$ K, respectively.  They are consistent with those of previous observations (e.g. \cite{Goss1985}, \cite{Mills}),  
although the typical uncertainty is estimated to be as large as 10\% of the derived value. 
In addition,  the measured line brightness temperatures (see Figure 3) are much smaller than the derived electron temperatures.  These indicate that  the optically thin assumption is valid in the compact H$_{\mathrm{II}}$ regions.
The mean electron temperatures are also summarized in Table 1.

\begin{figure}
\begin{center}
\includegraphics[width=10cm, bb=0 0 306.9 362.29]{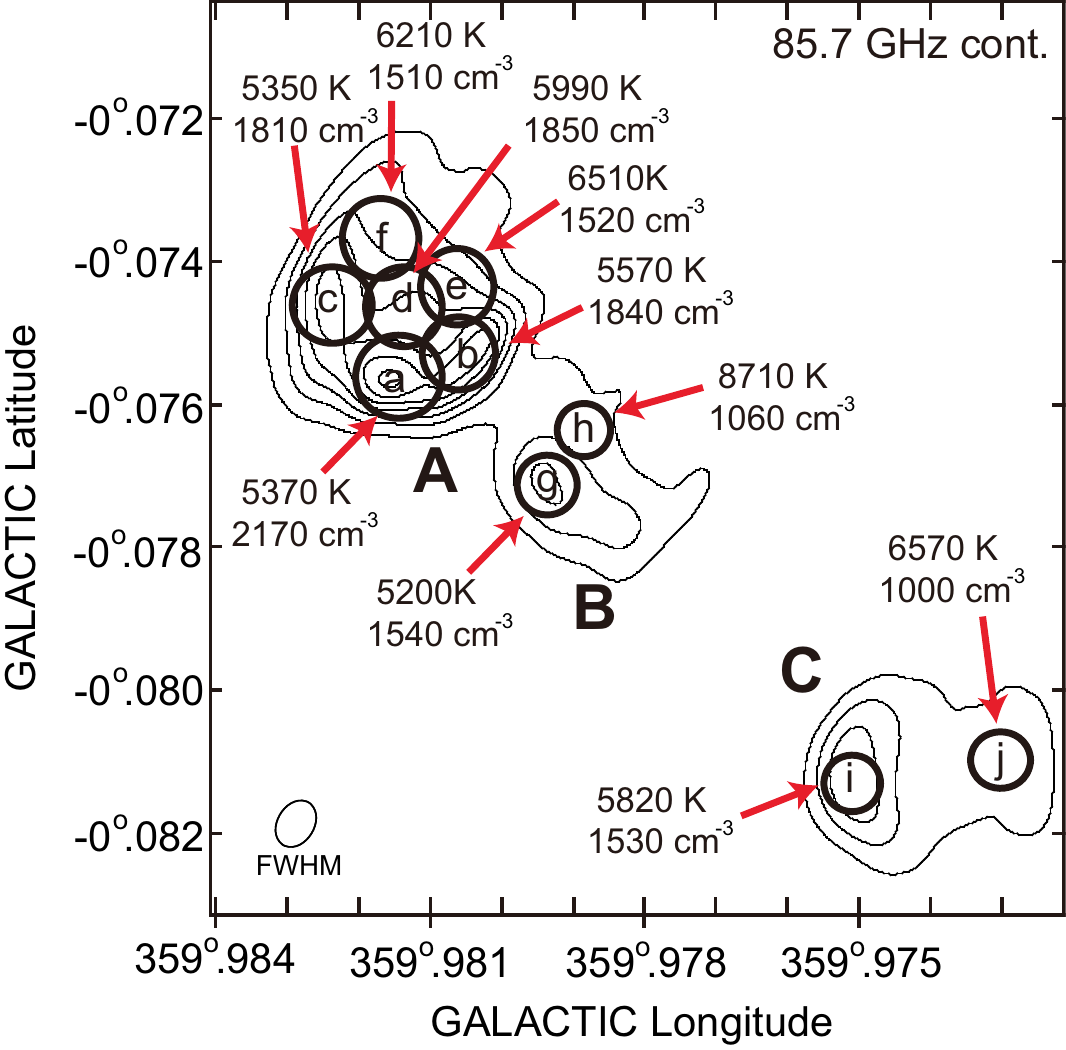}
\end{center}
\caption{The electron temperatures and electron densities at typical positions in the compact H$_{\mathrm{II}}$ regions in G-0.02-0.07, overlaid on the continuum map at 85.7 GHz. The integration areas are shown as circles.  }
\end{figure}

\begin{table}
  \caption{Physical properties at typical positions in G-0.02-0.07. }
  \label{tab:first}
  \begin{center}
    \begin{tabular}{cccccccc}
    \hline
    \hline
Region$^1$ &$l$&$b$&$T^{\ast ~2}_{\mathrm e}$&$n^{~~2}_{\mathrm e}$& $C_\mathrm{s}^{~~2,3}$& $V_{\mathrm{exp}}^{~~~2,4}$&Remarks\\
&[deg.]&[deg.]&[K]& [cm$^{-3}$]&[km s$^{-1}$]&[km  s$^{-1}$]&\\
\hline
a&359.9816&-0.0756&5370&2170&9.5&11.7&bright-half of HII-A\\
b&359.9804&-0.0753&5570&1840&9.7 &13.0&bright-half of HII-A \\
c&359.9824&-0.0745&5350&1810&9.5&12.3&bright-half of HII-A\\
d&359.9815&-0.0744&5990&1850&10.1 &12.1&dark-half of HII-A\\
e&359.9801&-0.0743&6510&1520&10.5 &7.2&dark-half of HII-A\\
f&359.9816&-0.0733&6210&1510 &10.2&10.1&dark-half of HII-A\\
g&359.9793&-0.0772&5200&1540 &9.4 &12.1&bright-half  of HII-B\\
h&359.9784&-0.0760&8710&1060& 12.1&9.5&dark-half of HII-B\\
i&359.9750&-0.0812&5820&1530 &9.9 &9.3&bright-half of HII-C\\
j&359.9727&-0.0809&6570&1000& 10.5&11.9&dark-half of HII-C\\
 \hline
    \end{tabular}
 \end{center}
$^1$ The regions are shown in Figure 9.
$^2$~The typical uncertainty is estimated to be as large as 10\% of the derived value.
$^3$~Sound velocity, $C_\mathrm{s}=(2kT^\ast_\mathrm{e}/m_\mathrm{H})^{0.5}$.
$^4$~Expanding velocity, $V_{\mathrm{exp}}=\frac{1}{2}(\Delta V_{\mathrm{FWHM}}^2-(2\sqrt{\ln2}C_\mathrm{s})^2)^{0.5}$.
\clearpage
\end{table}
Using the equation (2),  the electron temperatures at typical positions in HII-A are derived from the ratio between the 
integrated $\mathrm{H}42\alpha$ recombination line intensity and continuum flux density at 86 GHz. They are shown in Figure 9. The first three sampling positions (a, b, and c) correspond to the half-shell like brightening limb and the remaining three (d, e, and f) are located in the dimming half. 
The electron temperatures at a, b, and c are estimated to be $T^\ast_{\mathrm e}=5370$, $5570$, and $5350$ K, respectively. Meanwhile, the electron temperatures at d, e, and f are estimated to be $T^\ast_{\mathrm e}= 5990$, $6510$, and $6210$ K, respectively. Moreover, the estimated electron temperatures on the half-shell like brightening limbs and the dimming halves are $T^\ast_{\mathrm e}= 5200$ and $8710$ K in HII-B and $T^\ast_{\mathrm e}= 5820$ and $6570$ K in HII-C, respectively.  The electron temperatures on the brightening  limb are slightly lower than those on the dimming limb.
These electron temperatures are  summarized in Table 2.

\subsubsection{Electron Density}
The  electron density, $n_{\mathrm e}$, in the compact H$_{\mathrm{II}}$ regions is estimated from the continuum brightness temperature, $T_{\mathrm B}$,
and  the electron temperature derived above, $T^\ast_{\mathrm e}$, and the path length of the ionized gas, $L$, assuming that the continuum emission is optically thin. 
The well-known formula of the electron density is given by
\begin{equation}
\label{3}
n_{\mathrm e}[{\mathrm cm}^{-3}]=\left[\frac{T_{\mathrm B}T_{\mathrm e}^{\ast 0.35}\Big(\frac{\nu}{\mathrm{GHz}}\Big)^{2.1}}{8.235\times10^{-2}\alpha(\nu,T)\Big(\frac{L}{\mathrm{pc}}\Big)}\right]^{0.5}
\end{equation}
\citep{Altenhoff}. We assume here that the ionized gases have  spherical  shapes with the radius of $R$ and constant electron density of $\bar{n_{\mathrm e}}$. The mean path length is given by $L=\frac{4\pi}{3}R^3/\pi R^2=\frac{4}{3}R$. Here the $R$ is assumed to be  $R=\bar{r}$ in 3.1.  The electron densities of HII-A, HII-B, HII-C, and HII-D are estimated to be $\bar{n_{\mathrm e}}=1300$, $1000$, $950$, and $2340$ cm$^{-3}$, respectively. These values may be somewhat  lower than those of previous observations (e.g. \cite{Mills}) although these are still  consistent with typical values in  the HII regions in the Galaxy. 
In addition, the mean emission measures,  $\bar{EM}=(4/3)n_{\mathrm e}^2R_\mathrm{s}$, are also estimated. These  physical parameters are also consistent with previous estimates although the typical uncertainty of the derived values is estimated to be as large as 10\% of the value. These are also summarized in Table 1.

The electron densities  at typical positions shown in Figure 9 are also derived using the same assumptions.
The electron densities at a, b, and c are estimated to be $n_{\mathrm e}=2170$, $1840$, and $1810$ cm$^{-3}$, respectively. Meanwhile, the electron densities at d, e, and f are estimated to be $n_{\mathrm e}= 1850$, $1520$, and $1510$ cm$^{-3}$, respectively. Moreover, the estimated electron densities on the half-shell like limbs and the dark parts are $n_{\mathrm e}= 1540$ and $1060$ cm$^{-3}$ in HII-B and $n_{\mathrm e}= 1530$ and $1000$ cm$^{-3}$ in HII-C, respectively.  The electron densities on the brightening halves are slightly higher than those on the dimming halves.
These electron densities are also summarized in Table 2.
\begin{figure}
\begin{center}
\includegraphics[width=16cm, bb=0 0 1152.68 946.43]{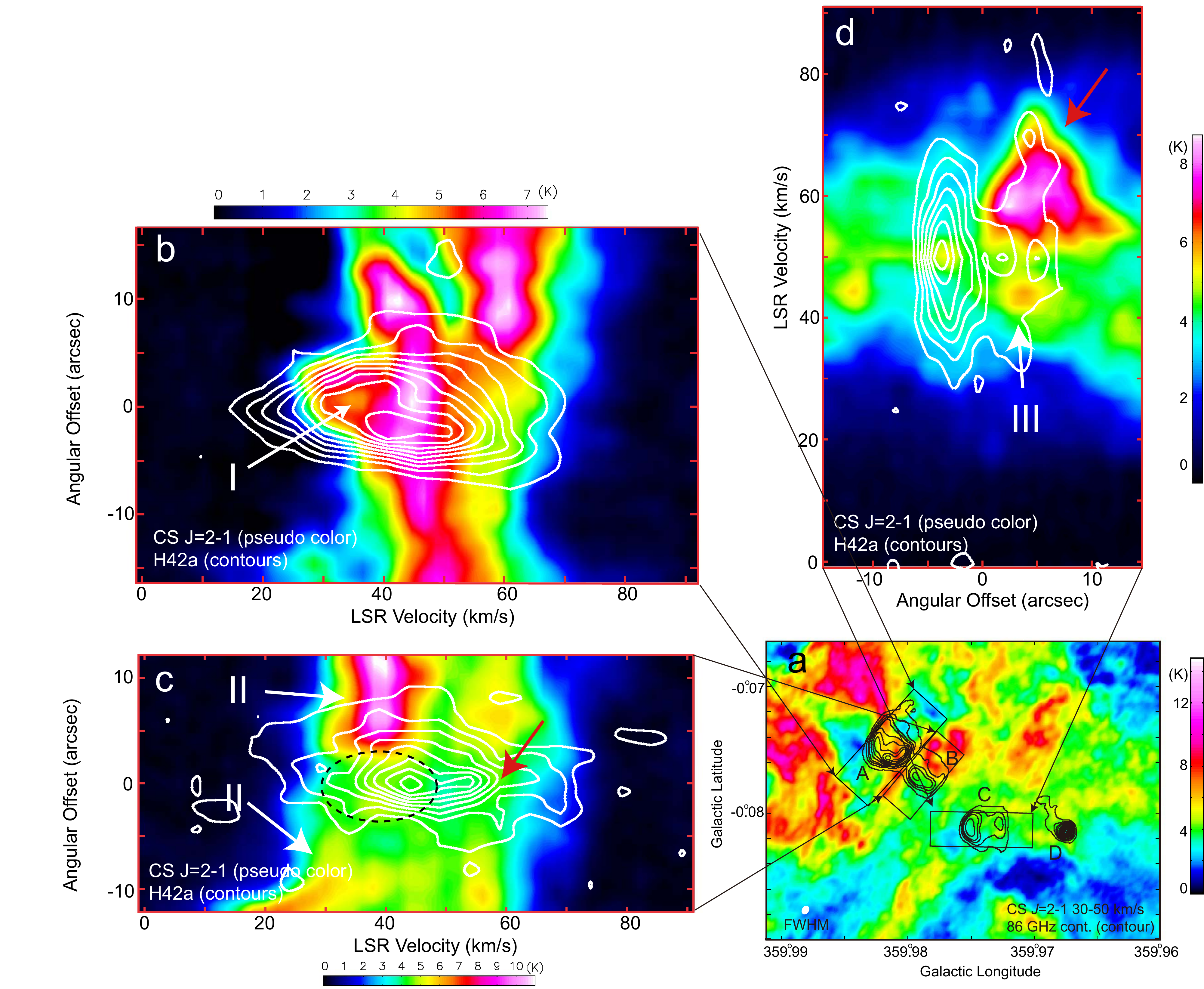}
\end{center}
\caption{
{\bf a} Comparison between continuum map of HII-A, HII-B, and HII-C at 86 GHz (contours)  and the integrated intensity maps of the CS $J=2-1$ emission line  from $V_{\mathrm{LSR}}=30$ to $50$ km s$^{-1}$ (pseudo color). The angular resolution is $1.9\arcsec \times 1.3\arcsec$($PA=-37^\circ$) shown on the lower-left corner as an oval. The contour levels are the same as those in Figure 1.  
 {\bf b} Position-velocity (PV) diagram of the CS $J=2-1$ emission line (pseudo color) and H42$\alpha$ recombination line (contours) along a rectangle including HII-A shown in {\bf a}. The first contour level and interval are both $0.072$ K in $T_{\mathrm{B}}$.  The Greek numbers in the PV diagrams also indicates molecular filaments  shown in Figure 4 and Figure 6. 
 {\bf c} PV diagram  of HII-B along a rectangle shown in {\bf a}. The first contour level and interval are  both $0.043$ K in $T_{\mathrm{B}}$.
  {\bf d} PV diagram of HII-C along a rectangle shown in {\bf a}. The first contour level and interval are both $0.072$ K in $T_{\mathrm{B}}$.} 
\end{figure}

\begin{figure}
\begin{center}
\includegraphics[width=16cm, bb=0 0 1152.63 950.72]{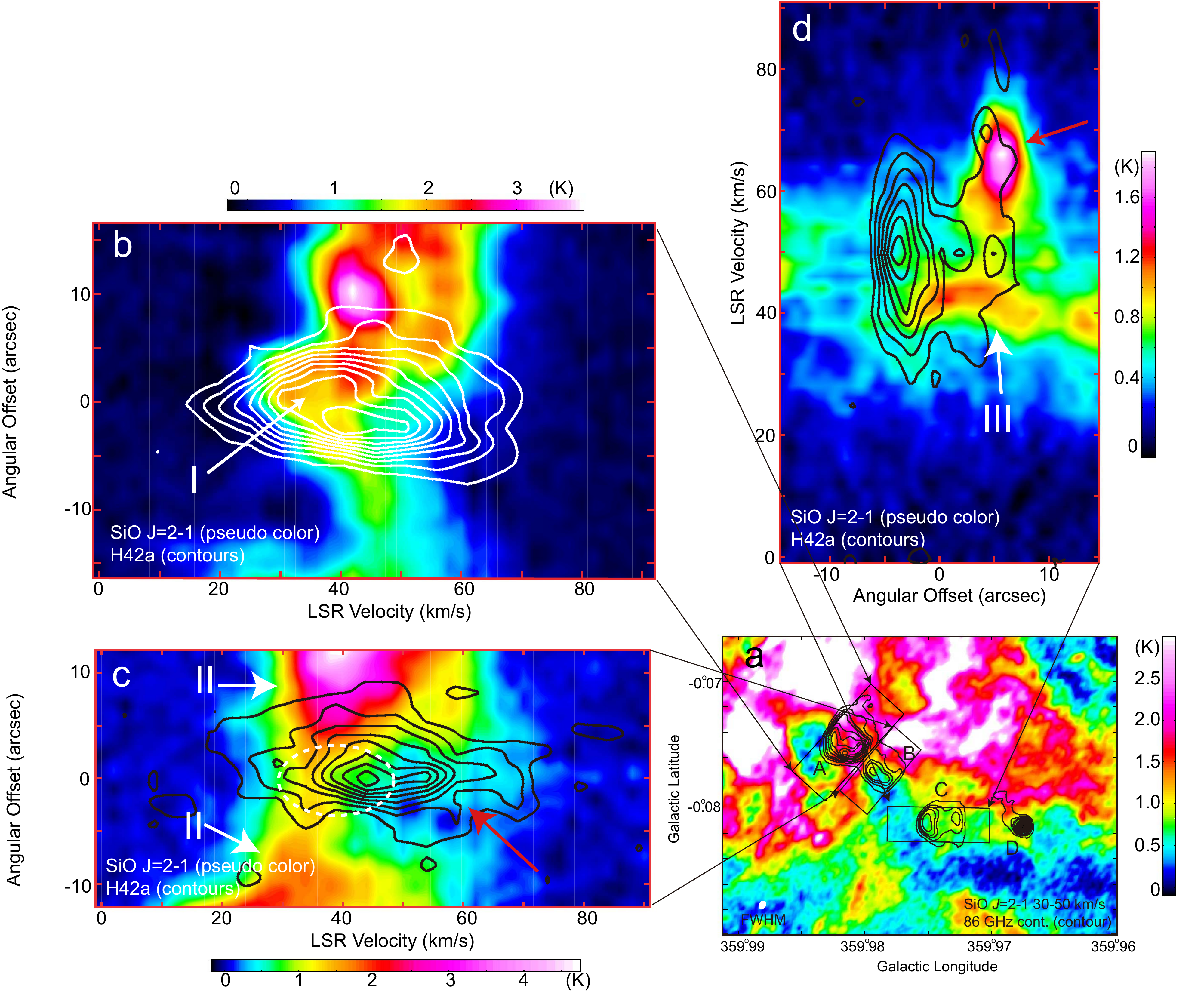}
\end{center}
\caption{
{\bf a} Comparison between continuum map of HII-A, HII-B, and HII-C at 86 GHz (contours)  and the integrated intensity maps of the SiO $J=2-1$ emission line  from $V_{\mathrm{LSR}}=30$ to $50$ km s$^{-1}$ (pseudo color). The angular resolution is $2.0\arcsec \times 1.4\arcsec$($PA=-40^\circ$) shown on the lower-left corner as an oval. The contour levels are the same as those in Figure 1.  
 {\bf b} Position-velocity (PV) diagram of the SiO $J=2-1$ emission line (pseudo color) and H42$\alpha$ recombination line (contours) along a rectangle including HII-A shown in {\bf a}. The first contour level and interval are both $0.072$ K in $T_{\mathrm{B}}$.  The Greek numbers in the PV diagrams also indicates molecular filaments  shown in Figure 4 and Figure 6. 
 {\bf c} PV diagram  of HII-B along a rectangle shown in {\bf a}. The first contour level and interval are  both $0.043$ K in $T_{\mathrm{B}}$.
  {\bf d} PV diagram of HII-C along a rectangle shown in {\bf a}. The first contour level and interval are both $0.072$ K in $T_{\mathrm{B}}$.} 
\end{figure}

\subsection{Relation of the H$_{\mathrm{II}}$ regions A-C with Molecular Gas}
Figure 10a and Figure 11a show the molecular gas  observed in the CS $J=2-1$ and SiO $J=2-1$ emission lines, respectively. The integrated velocity ranges  of these maps are both from $V_{\mathrm{LSR}}=30$ to $50$ km s$^{-1}$. As mentioned previously, we use here the data of these emission line after missing-flux compensation with the single-dish data combining \citep{Uehara} to depict the widely-extended ambient molecular gas  in the 50MC.  These maps also show the H$_{\mathrm{II}}$ regions observed in the 86 GHz continuum emission (contours).  
The H$_{\mathrm{II}}$ regions A-C seem to correspond to dimming parts of the molecular gas filaments which are embedded  in the extended ambient molecular gas as mentioned in the previous section (also see Figure 4 and Figure 6).

Figure 10b and Figure 11b show the PV diagrams around HII-A  in the CS $J=2-1$ and SiO $J=2-1$ emission lines, respectively. The sampling areas are shown as rectangles in the integrated velocity maps. The areas are crossing the molecular gas filament I. These figures also show the PV diagrams in the H42$\alpha$ recombination line for comparison (contours). The Greek numbers in the PV diagrams also indicates the molecular filaments  shown in Figure 4 and Figure 6.  In Figure 10b, the molecular gas filament I is identified as a shell-like component which is extending from $-5\arcsec$ to $5\arcsec$ in angular offset and from $30$ to $50$ km s$^{-1}$ in radial velocity.  The smaller velocity part of HII-A seems to fit in the shell-like component.
The component is also prominent  in the SiO $J=2-1$ emission line (Figure 11b). Because SiO molecules are enhanced by a shock wave, this suggests that there is shocked molecular gas associated with the part of HII-A. 

Figure 10c and Figure 11c show the PV diagrams around HII-B along the filament II in the CS $J=2-1$ and SiO $J=2-1$ emission lines, respectively.  The filament II is seen as a molecular gas ridge around $V_{\mathrm{LSR}}=20-50$ km s$^{-1}$ in these PV diagrams. 
There is a dip of the ridge around $-3\arcsec$ to $3\arcsec$  in  the CS $J=2-1$ emission line (a broken oval), which corresponds to the smaller velocity part of HII-B. While this part of the continuous ridge is also identified  in the SiO $J=2-1$ emission line (a broken oval), suggesting that SiO molecules are increased by the shock wave around the part of HII-B as in the HII-A case. 
Another gas ridge is identified around $V_{\mathrm{LSR}}=50-60$ km s$^{-1}$ in the CS $J=2-1$  emission line. The ridge also has a shallow dip toward HII-B (a red arrow). 
However, this ridge is not clear in the SiO $J=2-1$ emission line (a red arrow). 
 
Figure 10d and Figure 11d show the PV diagrams around HII-C in the CS $J=2-1$ and SiO $J=2-1$ emission lines, respectively.  
The filament III is seen as a molecular gas component extending from $-3\arcsec$ to $10\arcsec$ in angular offset and from $40$ to $50$ km s$^{-1}$ in radial velocity. The part corresponding to HII-C in the SiO $J=2-1$ emission line is more prominent than that in the CS $J=2-1$ emission line, suggesting that SiO molecules are also increased by the shock wave around the part of HII-C. 
In Figure 10d, another molecular gas component is identified around $V_{\mathrm{LSR}}=50-70$ km s$^{-1}$ and  from $0\arcsec$ to $13\arcsec$ in angular offset (a red arrow).  The negative angular offset side of the molecular gas component seems to be associated with the  ionized gas of HII-C. While the counterpart in the SiO $J=2-1$ emission line is identified there (a red arrow). They also suggest that the shock wave around HII-C enhances the SiO emission line partly.

Consequently, we interpret the physical relation between the  H$_{\mathrm{II}}$ regions and the molecular filaments as follows.
Lyman continuum photons emitting from embedded massive stars are eroding the molecular gas and the ionized gas fills the newly-made vacant spaces. 
As shown above, HII-A, HII-B and HII-C  seem to be located on the dimming parts of the molecular gas filaments in the $l-b-v$ space. The mean width of the molecular gas filaments found in the 50MC is  $W=0.27\pm0.06$ pc  (\cite{Uehara2017}, \cite{Uehara-b}).  While the mean diameters of the  H$_{\mathrm{II}}$ regions (see Table 1) are larger than the observed widths, $2\bar{r}> W$. 
The H$_{\mathrm{II}}$ regions have already broken through the molecular filaments and are growing in the ambient molecular gas although they may be still involved partly in the filaments.
The surrounding gas of these H$_{\mathrm{II}}$ regions is detected in the CS $J=2-1$ emission line (also see Figure 4). There is a good complementary relation  between the ionized and molecular gas. Meanwhile the relation is not so good for the molecular gas detected in the H$^{13}$CO$^+$ $J=1-0$ emission line  (see Figure 5). 
There are the components including shocked molecular gas around  these H$_{\mathrm{II}}$ regions. However, the H$_{\mathrm{II}}$ regions are not totally surrounded by such components. 
The shock wave would be suppressed by any reason or SiO molecules may have become extinct partly even in the shocked molecular gas.  SiO molecules can be dissociated by softer UV photons escaped from the H$_{\mathrm{II}}$ regions.
The photodissociation rate for FUV field of SiO molecules in molecular clouds is fairly larger than that of CS molecules \citep{Martin2012}.

\subsection{Evolution of H$_{\mathrm{II}}$ Regions}
First we assume here the picture that the H$_{\mathrm{II}}$ regions evolve by the I-front eroding  the  ambient  molecular gas.
According to the classical recipe of such a case (\cite{Spitzer},  \cite{Draine2011}),  the propagation of the I-front is given by
\begin{equation}
\label{4 }
4\pi R_{\mathrm{i}}^2n_0dR_{\mathrm{i}}=\Big(Q_0-\frac{4\pi}{3}R_{\mathrm{i}}^3\alpha_{\mathrm{B}}n_0^2\Big)dt,
\end{equation}
where $n_0$ is the Hydrogen molecule gas density of the  ambient molecular gas and $\alpha_{\mathrm{B}}$ is the Case B recombination coefficient. The  Case B recombination coefficient is approximately given by 
\begin{equation}
\label{5}
 \alpha_{\mathrm{B}}[\mathrm{cm}^3 \mathrm{s}^{-1}]\simeq2.56\times10^{-13}[T^\ast_\mathrm{e}/10^4 \mathrm{K}]^{-0.83},
 \end{equation}
 (\cite{Draine2011}).
We also assume that the emitting rate of hydrogen-ionizing photons from the central star is a constant, $Q_0$ ($Q_0\sim1\times10^{48}$ s$^{-1}$ for late O stars; e.g. \cite{Martins2005}), although this may be a variable at the early evolutionary stage.

\begin{table}

  \caption{The calculated brightness temperature of the CS $J=2-1$ emission line }
  \label{tab:first}
 \begin{center}
    \begin{tabular}{cccc}
    \hline
    \hline
Ambient molecular gas, $n_0$ 
&$T_{\mathrm{B}}$$^1$ at $T_{\mathrm{K}}=100$K&$T_{\mathrm{B}}$$^1$ at $T_{\mathrm{K}}=150$K&$T_{\mathrm{B}}$$^1$ at $T_{\mathrm{K}}=200$K\\
$[\mathrm{cm}^{-3}]$& [K] & [K]& [K]\\
\hline
2500&0.3&0.4&0.5\\
5000&1.1&1.3&1.6\\
7500&2.1&2.5&2.8\\
10000&3.1&3.7&4.2\\
12500&4.2&5.0&5.7\\
 15000&5.3&6.3&7.1\\
17500&6.4&7.6&8.6\\
20000&7.5&8.9&10.0\\
22500&8.6&10.2&11.4\\
25000&9.7&11.5&12.9\\
 \hline
   \end{tabular}
 \end{center}
$^1$ The values are based on the non-LTE calculation, RADEX (\cite{VanderTak}). 
The velocity width of the CS $J=2-1$ emission line is assumed to be $\Delta V=15$ km s$^{-1}$.
The fractional abundance of CS molecule is also assumed to be $X{\mathrm{(CS)}}=1\times 10^{-8}$.
\clearpage
\end{table}
We would derive the gas density of the  ambient  molecular gas, $n_0$, using the non-LTE calculation, RADEX (\cite{VanderTak}).
The typical brightness temperature and velocity width of the CS $J=2-1$ emission line are observed to be $T_{\mathrm{B}}\sim 6$ K and $\Delta V=15$ km s$^{-1}$, respectively (see Figure 10).  The path length is assumed to be  $l \sim 3$ pc from the diameter of the  50MC. The fractional abundance and gas kinetic temperature of CS molecule are assumed to be $X{\mathrm{(CS)}}=1\times 10^{-8}$, which is the typical value in the Galactic disk region,  and $T_{\mathrm{K}}=150$ K (e.g \cite{Uehara}), respectively.
The brightness temperatures are calculated for $n_0$ as a parameter. They are summarized in Table 3.
Comparing the observed brightness temperature with the calculated values, the  gas density  is estimated to be $n_0\sim 1.5\times10^4$ cm$^{-3}$. The uncertainty is guessed to be as large as  $\pm50$ \%.

Integrating the equation (4), the radius of the H$_{\mathrm{II}}$ region, $R_{\mathrm{i}}$, and the advancing velocity of the I-front, $V_{\mathrm{i}}$, are given by
\begin{equation}
\label{6}
R_{\mathrm{i}}=R_{\mathrm{S}}\Big(1-e^{-tn_0\alpha_{\mathrm{B}}}\Big)^{1/3},
\end{equation}
\begin{equation}
\label{7}
V_{\mathrm{i}}=\frac{dR_{\mathrm{i}}}{dt}=\Big(\frac{Q_0n_0\alpha_{\mathrm{B}}^2}{36\pi}\Big)^{1/3}\frac{e^{-tn_0\alpha_{\mathrm{B}}}}{\Big(1-e^{-tn_0\alpha_{\mathrm{B}}}\Big)^{2/3}},
\end{equation}
where $R_{\mathrm{S}}$ is the so-called Str\"omgren radius, $R_{\mathrm{S}}=\Big(\frac{3Q_0}{4\pi\alpha_{\mathrm{B}}n_0^2}\Big)^{1/3}$, 
(\cite{Spitzer}, \cite{Draine2011}).
In addition,  the sound  velocity  in the Hydrogen ionized gas is given by 
\begin{equation}
\label{8}
C_\mathrm{s}=(2kT_\mathrm{i}/m_\mathrm{H})^{0.5}=13\times[T^\ast_\mathrm{e}/10^4\mathrm{K}]^{0.5}
 [\mathrm{km~s}^{-1}]
  \end{equation}
assuming that the ion temperature, $T_\mathrm{i}$, is equal to the LTE electron temperature, $T^\ast_\mathrm{e}$. 

In the case that the electron temperature is $T^\ast_\mathrm{e}=6000$ K, the ambient molecular gas density is $n_0= 1.5\times10^4$ cm$^{-3}$, and the emitting rate of hydrogen-ionizing photons is $Q_0=1\times10^{48}$ s$^{-1}$,  the advancing velocity of the I-front decreases to the sound  velocity  in the Hydrogen ionized gas  at $t_\mathrm{0}\simeq 30$ years; $V_{\mathrm{i}}\simeq C_\mathrm{s}$.  The radius of the I-front has increased to $R_{\mathrm{i}}\simeq R_{\mathrm{S}}$ at the time, $t=t_\mathrm{0}$.  The calculations are shown  in Figure 12 (Cf. \cite{Draine2011}).
The  I-front  advances into the ambient  molecular  gas without shock wave (so-called R-type I-front) when the advancing velocity is higher than the sound  velocity, i.e. $V_{\mathrm{i}}> C_\mathrm{s}$ (\cite{Spitzer}, \cite{Draine2011}).    After the advancing velocity decreases to the sound  velocity, i.e. $V_{\mathrm{i}}\simeq C_\mathrm{s}$,  the shock wave takes place in front of the I-front and compresses the ambient  molecular  gas  (so-called D-type I-front) (\cite{Spitzer}, \cite{Draine2011}). 
However,  SiO emission components adjacent to the  H$_{\mathrm{II}}$ regions are detected but not fully surrounding them as shown in the previous subsection.  The SiO molecule abundance may become lower than the detection limit because the SiO molecules are easily photodissociated by UV photons.

The gas pressure ratio of the observed ionized gas and the observed ambient molecular gas is estimated to be 
\begin{equation}
\label{9}
\frac{P_{\mathrm{HII} }}{P_{\mathrm{MC}}}= \frac{2n_\mathrm{e}kT^\ast_\mathrm{e}}{n_\mathrm{0}kT^\ast_\mathrm{MC,K}}.
 \end{equation}
When the radius of the ionized gas (R-type I-front) reached around the Str\"omgren radius, the gas pressure of the ionized gas was much higher than that of  the ambient molecular gas because the electron densities should be similar to the number density of the ambient molecular gas, $n_\mathrm{e}\simeq n_\mathrm{0}$. 
The observed gas pressure of the ionized gas is found to be higher than the observed gas pressure of the ambient molecular gas now. Therefore the expanding velocity of  the ionized gas should be as large as  the sound  velocity  in the ionized gas. 
From the  derived electron temperatures  (see Table 1), the mean sound  velocities in HII-A, HII-B, HII-C, and HII-D are estimated to be $C_{\mathrm{s}}=9.9, 9.3, 10.0$, and $9.6$ km$s^{-1}$, respectively. 
The  mean expanding velocity of the ionized gas, $V_{\mathrm{exp}}$, would be estimated from the observed FWHM velocity width, $\Delta V_{\mathrm{FWHM}}$, of the line profiles in the $\mathrm{H}42\alpha$ recombination line (see Figure 4) and the sound velocities derived above using the formula  given by 
\begin{equation}
\label{10}
V_{\mathrm{exp}}\simeq\frac{1}{2}(\Delta V_{\mathrm{FWHM}}^2-(2\sqrt{\ln2}C_\mathrm{s})^2)^{0.5}.
 \end{equation}
Here the Maxwell distribution is assumed for convenience. Then the mean expanding velocities of HII-A, HII-B, HII-C, and HII-D are $V_{\mathrm{exp}}=16.7, 11.6, 11.1$, and $12.1$ km s$^{-1}$, respectively.
The mean expanding velocities seem to be slightly larger than the sound velocities, $V_{\mathrm{exp}}\gtrsim C_\mathrm{s}$. 
These indicate that the I-fronts of HII-A, HII-B, HII-C, and HII-D are now expanding with the nearly sound velocities as expected above. The typical evolution of the radius of the ionized gas is also shown in Figure 12. 

On the other hand, the derived electron densities are lower than the number density of the ambient molecular gas now.  The decrease of the electron density was probably caused by the expansion of the ionized gas. 
The relation between the present radius of the ionized gas, $\bar{r}$,  and the Str\"omgren radius is given by 
\begin{equation}
\label{11}
\bar{r}\sim \Big(\frac{n_\mathrm{0}}{n_\mathrm{e}}\Big)^{2/3} R_\mathrm{s}
 \end{equation}
Using this formula, the Str\"omgren radii of HII-A, HII-B, and HII-C are estimated to be $0.06, 0.04$, and $0.04$ pc, respectively. 
 Therefore the  elapsed time, $t_\mathrm{1}$ , from the Str\"omgren radius to the present radius are given by
 \begin{equation}
\label{12}
t_\mathrm{1}\sim\frac{\bar{r}-R_\mathrm{s}}{V_{\mathrm{exp}}}.
 \end{equation}
Because the the elapsed time to the Str\"omgren radius is as short as $t_\mathrm{0}\sim 30$ years, the derived  elapsed times could be considered as the ages of the  H$_{\mathrm{II}}$ regions themselves, $t_\mathrm{age}\simeq t_\mathrm{1}$.  Using this formula, the ages of HII-A, HII-B, and HII-C are estimated to be $t_\mathrm{age}\simeq1.4\times10^4, 1.7\times10^4$, and $2.0\times10^4$ years, respectively. 
Although the typical uncertainty of the age is as large as $30~\%$, which would mainly come from the uncertainty of the ambient gas density, the  H$_{\mathrm{II}}$ regions would be formed within $\lesssim1\times10^4$ years. Moreover, they might be formed from southwest to northeast sequentially. In addition, the age of HII-D is estimated to be $t_\mathrm{age}\simeq0.7\times10^4$ years with the same procedure.
 
\begin{figure}
\begin{center}
\includegraphics[width=14cm, bb=0 0 471.96 581.8 ]{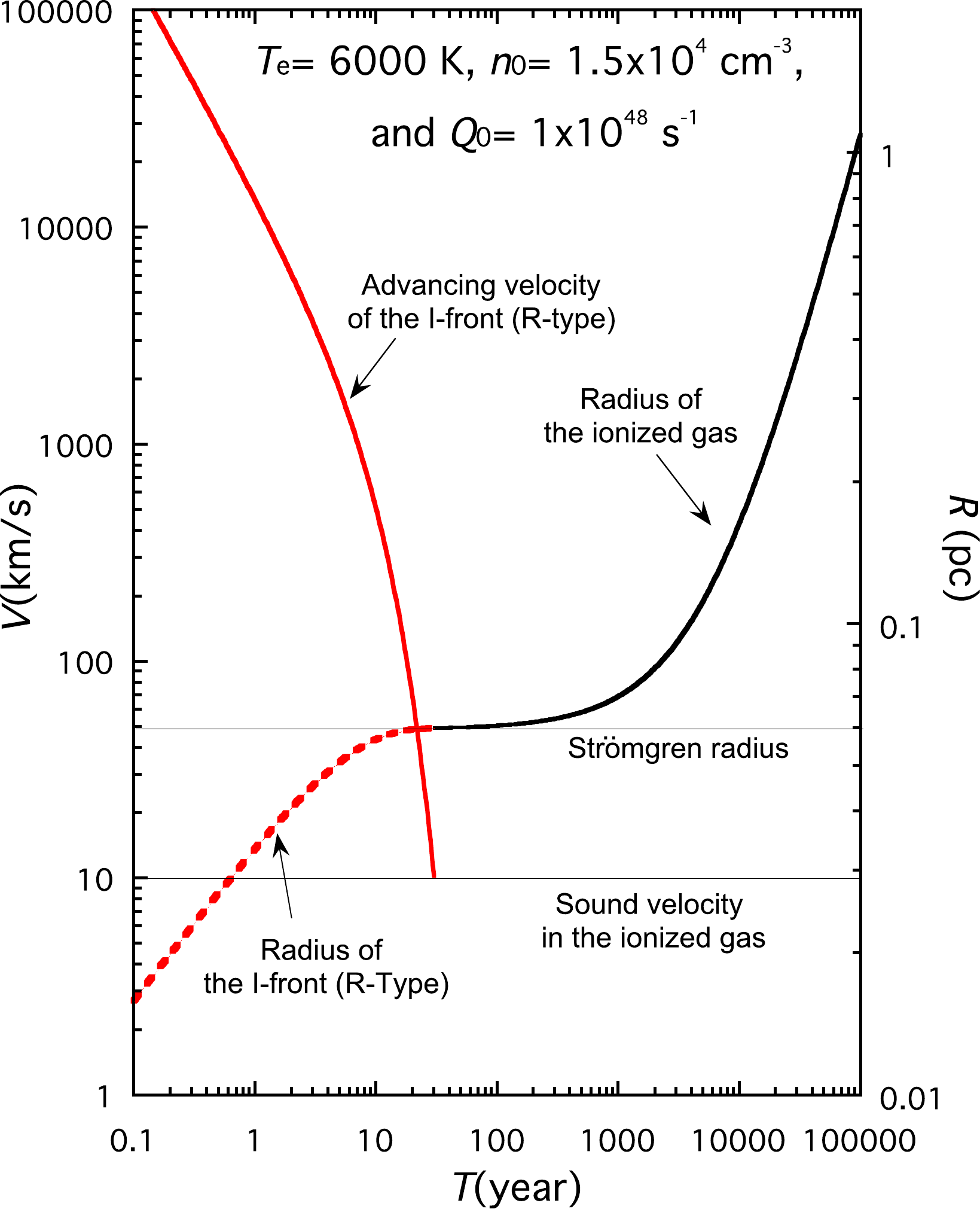}
\end{center}
\caption{Evolution of typical H$_{\mathrm{II}}$ region in the case that the electron temperature is $T^\ast_\mathrm{e}=6000$ K, the ambient number density is $n_0= 1.5\times10^4$ cm$^{-3}$, and the emitting rate of hydrogen-ionizing photons is $Q_0=1\times10^{48}$ s$^{-1}$.  
In the phase of R-type I-front, the radius and advancing velocity of the I-front are given by $R_{\mathrm{i}}=R_{\mathrm{S}}\Big(1-e^{-tn_0\alpha_{\mathrm{B}}}\Big)^{1/3}$,$
V_{\mathrm{i}}=\Big(\frac{Q_0n_0\alpha_{\mathrm{B}}^2}{36\pi}\Big)^{1/3}e^{-tn_0\alpha_{\mathrm{B}}}\Big(1-e^{-tn_0\alpha_{\mathrm{B}}}\Big)^{-2/3}$. The ionized gas (the D-type I-front) is assumed to be expanding with the sound velocity of the ionized gas.} 
\end{figure}

From the definition of the Str\"omgren radius, the emitting rate of hydrogen-ionizing photons, $Q_0$, is given by
\begin{equation}
\label{13}
 Q_0 = \frac{4\pi}{3}R_\mathrm{s}^3 \alpha_Bn_0^2.
 \end{equation}
As mentioned above, the Str\"omgren radii are estimated from the measured radii, electron density, and ambient density. The emitting rates of HII-A, HII-B, HII-C, and HII-D are estimated to be $Q_0 = 2.3\times10^{48}$, $0.8\times10^{48}$, $0.9\times10^{48}$, and $0.5\times10^{48}$ s$^{-1}$, respectively.
Assuming that  the central stars are main sequence single stars, the comparisons of the emitting rates of hydrogen-ionizing photons with the calculated calibration (\cite{Martins2005}) indicate that the spectral types of the central stars in HII-A, HII-B, HII-C, and HII-D are O8V, O9.5V, O9V, and B0V, respectively. These parameters  are also summarized in Table 1. 
These derived spectral types are roughly consistent with the previous radio estimation (O7V, O8.5V, O8.5V, and O9V: \cite{Mills}) and are slightly later  than the previous IR estimation (O6V, O6V, O5V, and O9V: \cite{Lau2014}). 
The Helium ionized photon emitting rate of HII-D is expected to be much smaller than those of other H$_{\mathrm{II}}$ regions. This is consistent with non-detection of the He42$\alpha$ recombination line toward HII-D (see Section 3.4).
The ambiguity of our estimation would be caused by our simple assumptions, for example "spherical ionized gas expanding in the ambient molecular gas with uniform density", and the uncertainties of the derived parameters of  $T_\mathrm{e}, n_\mathrm{e}, n_0,$ and $\bar{r}$.

\subsection{Inner structures of HII-A, HII-B, and HII-C}
HII-A, HII-B, and HII-C each are clearly resolved into a bright-half and a dark-half as mentioned in Sec.3.1. It is an open question what makes these conspicuous characteristics of the H$_{\mathrm{II}}$ regions. 
A possible cause of the asymmetrical brightness distributions is ``bow shock" originated in the H$_{\mathrm{II}}$ regions (\cite{Yusef-Zadeh2010}, \cite{Lau2014}).
If the H$_{\mathrm{II}}$ region itself is moving in the ambient molecular gas with a velocity larger than the sound velocity in the ionized gas, the front half can make a bow shock which would compress the ionized gas and the surrounding molecular gas. In this case, a compressed ionized gas should make the bright-half of the H$_{\mathrm{II}}$ region and a shocked  molecular gas shell should be observed in front of it.  

However, there is no evident sign of such shocked  molecular gas as shown in Figure 10 (also see Figure 4).
If anything, the shocked molecular gas  seems to be associated with the dark-halves not the bright-halves as shown in Figure 11. 
Because the electron densities in the bright-halves are slightly higher than those in the dark-halves and the electron temperatures in the bright-halves are slightly lower than those in the dark-halves(see Figure 9), there is no large difference between the gas pressures both in the halves.
In addition, the velocity centroid of the ionized gas is similar to that of the ambient molecular gas (see Figure 7), suggesting that the velocity difference between the ionized and molecular gases is not larger than the sound velocity in the ionized gas. 
Therefore there is no observational evidence supporting the bow shock picture.

\begin{figure}
\begin{center}
\includegraphics[width=18cm, bb=0 0 957.23 923.64]{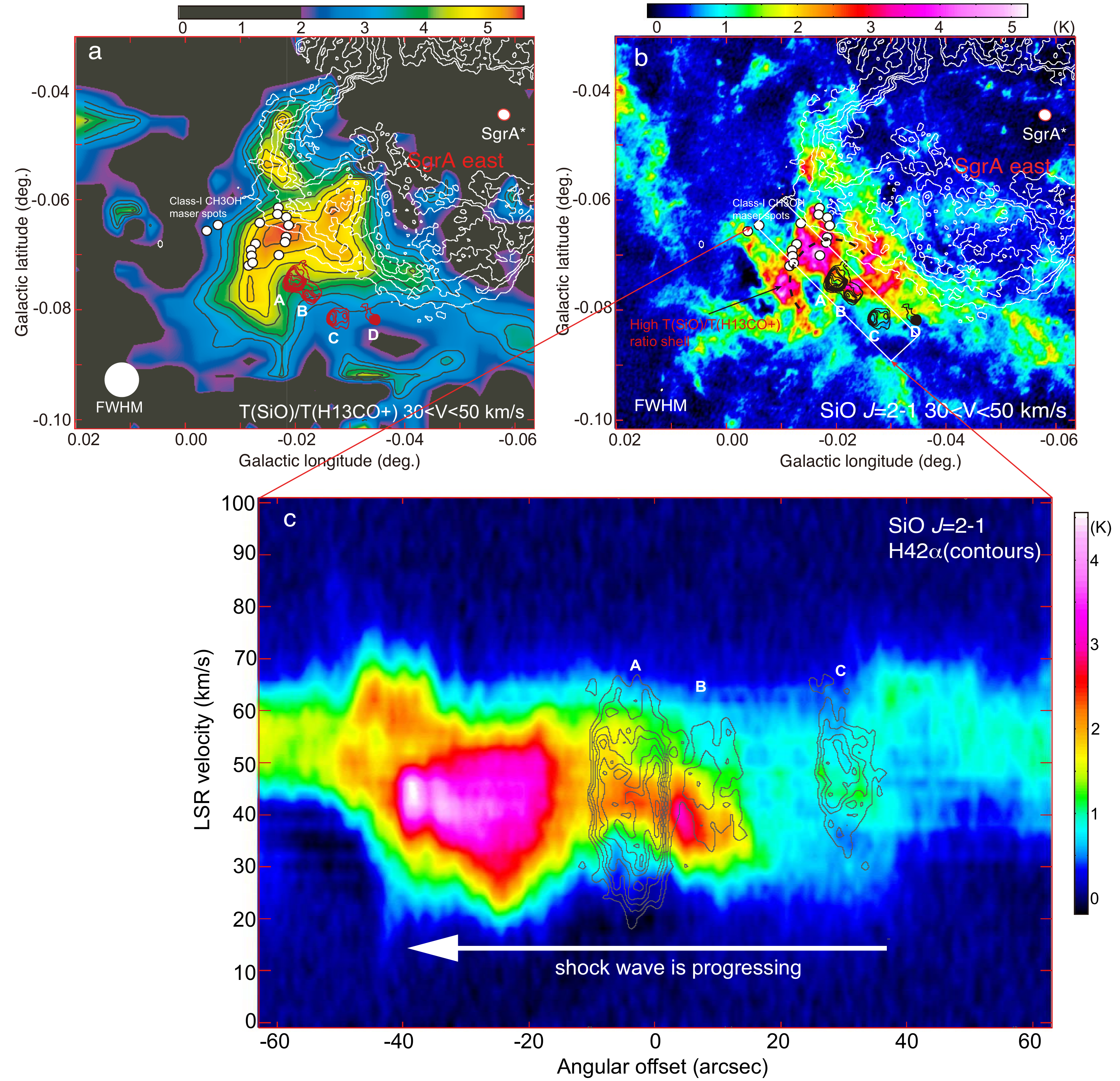}
\end{center}
\caption{Positional relation among the H$_{\mathrm{II}}$ regions, the SiO molecule enhancement area, and the Sgr A-E shell.  The H$_{\mathrm{II}}$ regions are shown by the 86-GHz continuum emission image which is the same as that in Figure 1 (contours).  The  Sgr A-E shell is shown by the 5-GHz continuum emission image with VLA (contours)  \citep{Yusef-Zadeh1987}. Class-I CH$_3$OH maser spots are also seen on the top of the feature (white filled circles) \citep{Phist}. 
{\bf a} The SiO molecule enhancement area  is shown as a half circular feature with high $T_\mathrm{B}\mathrm{(SiO)}/T_\mathrm{B}\mathrm{(H^{13}CO^+)}$ brightness temperature ratio ( \cite{Tsuboi2011}, \cite{Tsuboi2015}). The integration velocity range is $30<V<50$ km s$^{-1}$.  {\bf b}  The SiO molecule enhancement area is shown in the integrated intensity map of  the SiO emission line. The integration velocity range and mapping area are the same as those of {\bf a}.  A broken line curve indicates the ridge with $T_\mathrm{B}\mathrm{(SiO)}/T_\mathrm{B}\mathrm{(H^{13}CO^+)} >5$ shown in {\bf a}. {\bf c} PV diagrams of the SiO $J=2-1$ emission line and the H42$\alpha$ recombination line (contours) along a rectangle shown in {\bf b}. The first contour level and interval of the H42$\alpha$ recombination line are $0.1$ K and $0.05$ K in $T_{\mathrm{B}}$, respectively.} 
\end{figure}

\subsection{Origin of the Massive Young Stars in the H$_{\mathrm{II}}$ Regions}
As mentioned previously, the H$_{\mathrm{II}}$ regions are located in the molecular gas filaments and growing by ionizing  the ambient molecular gas.  These suggest that the massive young stars forming the H$_{\mathrm{II}}$ regions have been born in the molecular gas filaments in a similar way to the massive star formation in the Galactic disk region although even the star formation scenario has many issues yet. That is, massive molecular cloud cores were born in the filaments by any trigger, collapsed gravitationally, and formed massive stars. 

Figure 13a shows the previous observations of $T_\mathrm{B}\mathrm{(SiO)}/T_\mathrm{B}\mathrm{(H^{13}CO^+)}$ brightness temperature ratio in the 50MC using the Nobeyama 45-m telescope (\cite{Tsuboi2011}, \cite{Tsuboi2015}). The brightness temperature ratio is a good tracer of C-type shock wave propagating in the cloud because high ratio indicates the enhancement of SiO molecule.
In the figure, there are a half-circular feature with high ratio in the 50MC and another  feature with high ratio which traces the southeast limb of the SgrA East (SAE).  The half-circular feature would reflect a hollow hemisphere-like shocked molecular gas in the $l-b-v$ space \citep{Tsuboi2015}. Figure 13b shows the integrated intensity map using ALMA in the SiO emission line with the same area and velocity range as in the panel a.  The feature (broken line curve) is identified as a chain of the peaks in the SiO emission line map. HII-A, HII-B, and HII-C are located on the concave side of the feature. 

Figure 13c shows the PV diagrams of the SiO $J=2-1$ emission line and the H42$\alpha$ recombination line (contours) along a rectangle shown in Figure 13b. The half circular feature has a large velocity width of $\Delta V\gtrsim30$ km s$^{-1}$. This is consistent with that the C-type shock wave with the shock velocity exceeding $V_\mathrm{shock}\gtrsim30$ km s$^{-1}$ increases SiO molecules in the ambient molecular gas by sputtering of  dust grains efficiently (e.g. \cite{May2000, Gusdorf, Jim}).
The SiO $J=2-1$ emission line is decreasing with going to the positive direction.
Figures 13b also shows that the methanol maser spots in the velocity range of $V_\mathrm{LSR}=30-60$ km s $^{-1}$ are distributed on the north-east edge (or the top) of the feature (\cite{Phist}, see also Figure 3b  in \cite{Tsuboi2015}). The group of the maser spots are located around the extension of the line along HII-A, HII-B, and HII-C. Moreover, the spots corresponds to the negative-offset side of the SiO $J=2-1$ emission peak in Figure 13c.  Because the Class I methanol maser line at 44 GHz  is pumped  by the C-type shock wave and emitted immediately, the position of the present shock wave front  is expected to be around the group of the maser spots.  
Very high excited NH$_3$ emission lines up to $J=15$ have been detected toward the spots \citep{MillsMorris}. This indicates that the rotational temperature is higher than $T>400$ K, which may be caused by shock wave heating. These are consistent with the simulation of the gas temperature and chemical abundance (Cf. Figure 3 and Figure 4 in \cite{Gusdorf}) if the C-type shock wave propagated from southwest to northeast in the 50MC.
It has been reported that many  gravitationally-bound massive cores  have been located in the molecular gas filaments  in the 50MC \citep{Uehara}. 
Our interpretation is that 
the C-type shock wave propagated along the direction from HII-C to HII-A in the 50MC,
the shock wave would compress the molecular filaments, 
massive gravitationally bound  cores would be formed in the filaments by the  compression, 
massive star formation would begin in the cores, and  HII-C, HII-B and HII-A had been formed sequentially. 
Recent simulation studies also suggest that the observed PV diagrams are consistent with the C-type shock wave caused by a cloud-cloud collision in the 50MC (e.g. \cite{Haworth}). 

On the other hand, the SiO shocked feature located along the southeast limb of the SAE shell was probably caused by the shell because the shocked feature well traces the outside of the shell. The interaction between the 50MC and SAE had been discussed as a possible trigger for the massive star formation (e.g \cite{Yusef-Zadeh2010}). However, the H$_{\mathrm{II}}$ regions are located on the outside of the shocked molecular gas. The C-type shock wave caused by the expanding SAE shell does not yet reach the H$_{\mathrm{II}}$ regions.

\section{Summary}
We have observed the compact H$_{\mathrm{II}}$ region complex nearest  to the dynamical center of the Galaxy, G-0.02-0.07,  using ALMA  in the H42$\alpha$ recombination line,  CS $J=2-1$, H$^{13}$CO$^+ J=1-0$, and SiO $v=0, ~J=2-1$ emission lines, and 86 GHz continuum emission. This  is a part of  the first large-scale mosaic observation in the Sgr A complex.   
\begin{itemize}

  \item The H$_{\mathrm{II}}$ regions HII-A to HII-C in the cluster are clearly resolved into a shell-like feature with a bright-half and a dark-half in the recombination line and continuum emission.  
 
  \item The analysis of the absorption features in the molecular emission lines  show that HII-A, B and C are located on the near side of the 50MC but HII-D is located on the far side.
 
  \item The ranges of the electron temperature and density are $T_{\mathrm{e}}=5150-5920$ K and $n_{\mathrm{e}}=950-2340$ cm$^{-3}$, respectively. The electron temperatures on the bright-half are slightly lower than those on the dark-half. While the electron densities on the bright-half are slightly higher than those on the  dark-half. 

 \item   The H$_{\mathrm{II}}$ regions are located on the molecular filaments in the 50MC. They have already broken through the filaments and are growing in the ambient molecular gas.    There are some components with  shocked molecular gas around the H$_{\mathrm{II}}$ regions. However, they are not fully surrounded by such components.  

 \item  From the line width of the H42$\alpha$ recombination line, the expansion velocities of HII-A, HII-B, HII-C, and HII-D are estimated to be $V_{\mathrm{exp}}=16.7$, $11.6$, $11.1$, and $12.1$ km s$^{-1}$, respectively.   These are similar to the sound velocities in the ionized gas.
 
  \item  The expansion timescales of HII-A, HII-B, HII-C, and HII-D are estimated to be  $t_{\mathrm{exp}}\sim1.4\times 10^4$, $1.7\times 10^4$, $2.0\times 10^4$, and $0.7\times 10^4$ yr, respectively.

\item The spectral types of the central stars from HII-A to HII-D are estimated to be O8V, O9.5V, O9V, and B0V, respectively. 
These derived spectral types are roughly consistent with the previous radio estimation and are slightly later  than those in the previous IR estimation.  

\item The positional relation among the H$_{\mathrm{II}}$ regions, the SiO molecule enhancement area, and the Class-I maser spots strongly suggests that the shock wave caused by a cloud-cloud collision propagated along the line from HII-C to HII-A in the 50MC.  The shock wave would trigger the massive star formation.
\end{itemize}

\begin{ack}  
We would like to thank Dr. Ryan Lau for constructive comments as the referee.  
This work is supported in part by the Grant-in-Aids from the Ministry of Eduction, Sports, Science and Technology (MEXT) of Japan, No.16K05308 and No.19K03939. This paper makes use of the following ALMA data:ADS/JAO.ALMA\#2012.1.00080.S.  ALMA is a partnership of ESO (representing its member states), NSF (USA) and NINS (Japan), together with NRC(Canada), NSC and ASIAA (Taiwan), and KASI (Republic of Korea), in cooperation with the Republic of Chile.
The National Radio Astronomy Observatory (NRAO) is a facility of the National Science Foundation (NSF) operated under cooperative agreement by Associated Universities, Inc (AUI).  The Joint ALMA Observatory is operated by ESO, AUI/NRAO and NAOJ. 

\end{ack}

\end{document}